\documentclass[]{aa}
\usepackage[varg]{txfonts}
\usepackage{graphicx}
\RequirePackage[colorlinks=true,linkcolor=blue,citecolor=blue,urlcolor=blue]{hyperref}

\newcommand{\dcdot}{\mathbin{%
    \nonscript\mspace{-\muexpr\medmuskip*2/3}%
    \cdot
    \nonscript\mspace{-\muexpr\medmuskip*2/3}%
  }%
}

\begin{document}

\title{Zooming in on the Circumgalactic Medium with GIBLE: \\Tracing the Origin and Evolution of Cold Clouds}

\author{Rahul Ramesh\inst{1}\thanks{E-mail: rahul.ramesh@stud.uni-heidelberg.de}
\and Dylan Nelson\inst{1}
\and Drummond Fielding\inst{2,3}
\and Marcus Br\"{u}ggen\inst{4}
}
\institute{
Universität Heidelberg, Zentrum für Astronomie, ITA, Albert-Ueberle-Str. 2, 69120 Heidelberg, Germany \label{1}
\and Center for Computational Astrophysics, Flatiron Institute, 162 Fifth Avenue, New York, NY 10010, USA \label{2}
\and Department of Astronomy, Cornell University, Ithaca, NY 14853, USA \label{3}
\and Hamburg University, Hamburger Sternwarte, Gojenbergsweg 112, 21029 Hamburg, Germany \label{4}
}

\date{}

\abstract{We use the GIBLE suite of cosmological zoom-in simulations of Milky Way-like galaxies with additional super-Lagrangian refinement in the circumgalactic medium (CGM) to quantify the origin and evolution of CGM cold gas clouds. The origin of $z$\,$=$\,$0$ clouds can be traced back to recent ($\lesssim$\,$2$\,Gyr) outflows from the central galaxy ($\sim$\,45\,$\%$), condensation out of the hot phase of the CGM in the same time frame ($\sim$\,45\,$\%$), and to a lesser degree to satellite galaxies ($\lesssim$\,5\,$\%$). We find that in-situ condensation results from rapid cooling around local over-densities primarily seeded by the dissolution of the previous generation of clouds into the hot halo. About $\lesssim$\,10\,$\%$ of the cloud population is long lived, with their progenitors having already assembled $\sim$\,$2$\,Gyr ago. Collective cloud-cloud dynamics are crucial to their evolution, with coalescence and fragmentation events occurring frequently ($\gtrsim$\,20\,Gyr$^{-1}$). These interactions are modulated by non-vanishing pressure imbalances between clouds and their interface layers. The gas content of clouds is in a constant state of flux, with clouds and their surroundings exchanging mass at a rate of \mbox{$\gtrsim$\,$10^3$\,M$_\odot$\,Myr$^{-1}$}, depending on cloud relative velocity and interface vorticity. Furthermore, we find that a net magnetic tension force acting against the density gradient is capable of inhibiting cloud-background mixing. Our results show that capturing the distinct origins of cool CGM clouds, together with their physical evolution, requires high-resolution, cosmological galaxy formation simulations with both stellar and supermassive black hole feedback-driven outflows.
}


\titlerunning{Origin and Evolution of Cold Clouds with GIBLE}
\authorrunning{R. Ramesh et al.}

\maketitle

% ------------------------------------------------------------------------------

\section{Introduction}

Observations as well as theoretical models suggest that galaxies are surrounded by a halo of diffuse gas extending out to approximately the virial radius. Often referred to as the circumgalactic medium (CGM), this reservoir is believed to play an important role in galaxy growth and evolution (see \citealt{donahue2022} for a recent review of the CGM).

The phase-structure of the CGM is complex -- while a large fraction of the volume is dominated by a virialized warm-hot component, it also contains denser, colder gas. The high-velocity clouds (HVCs) observed in the Milky Way \citep[e.g.][]{muller1963,wakker1997}, as well as analogs in external galaxies \citep[e.g.][]{gim2021,weng2022}, are common examples. Although such gas clouds were first detected several decades ago, most key questions surrounding their formation and evolution remain open.

In particular, there is no consensus as to how these clouds originate. While theory suggests that condensation of hot halo gas triggered by density perturbations may form dense pockets of cold gas \citep{binney2009, fraternali2015, nelson2020,faerman2024}, ram pressure stripping of satellite gas as they infall towards the centre of their host halo may also be a possible source \citep{mayer2006,rohr2023}. Clouds may also originate as a result of galactic-scale outflows driven from galaxies by stellar and active galactic nuclei (AGN) feedback \citep{voit2015b,thompson2016,ramesh2023b}. Cool CGM gas could also be inflowing as part of a galactic fountain cycle \citep{fraternali2006}, or due to filamentary cosmological accretion, particularly at high redshift \citep{keres2009,nelson2016,mandelker2020}.

The lifetimes of cool clouds are similarly unknown. Early `cloud crushing' simulations suggested that they are expected to be short-lived, either as a result of fragmentation \citep{mellema2002,zhang2017} or complete destruction \citep{klein1994,schneider2017}. However, more recent results suggest that their survivability may be enhanced by certain physical processes. As an example, the Kelvin-Helmholtz (KH) instability, driven by velocity shear as clouds pass through their ambient media, creates an interface layer of warm gas. Rapid cooling of this interface can then increase the cold mass of the cloud, effectively improving its lifetime \citep{scannapieco2015,gronke2018,fielding2020}.

Other non-thermal processes also play a role: magnetic fields may provide additional pressure support to clouds \citep{sharma2010,nelson2020,fielding2023}, suppress fluid instabilities through their associated magnetic tension \citep{ji2018,sparre2020}, influence their shape \citep{kwak2009} and kinematics \citep{banda2016,bruggen2023}, or boost the drag force they experience in the presence of a draped layer \citep{ramesh2024b}, thereby decreasing the time needed for them to begin comoving with their surroundings \citep{mccourt2015}. Similar to magnetic fields, cosmic rays may also contribute pressure support \citep{butsky2020,bruggen2020}.

While theoretical studies have greatly improved our understanding of cloud growth and survival, most are based on idealized (i.e. wind tunnel-type) simulations. While such setups are typically inexpensive from a computational standpoint, thus allowing one to realise higher numerical resolution and/or explore varying physics, they have key limitations. First, their outcomes are sensitive to the initial conditions, which are ad hoc. The region(s) of parameter space probed may not correspond to true clouds, and generalization is difficult \citep{jennings2022}. Second, these setups generally assume clouds to be uniform density spheres embedded in uniform background media,\footnote{Note, however, that some recent studies have also started exploring cases of non-uniformity in the properties of the cloud and/or the background \citep{jung2023}.} and hence lack the diversity and complexity of a realistic CGM. Lastly, since a pre-existing cloud is initialised at the start of the simulation, these setups cannot address the question of origin.

An alternative approach is fully cosmological simulations. These evolve halo gas over cosmic epochs and cosmic scales, and thus offer a better chance at self-consistently reproducing a realistic CGM, albeit typically at coarser numerical resolution with respect to their idealised counterparts. However, recent projects have attempted to take a step forward in narrowing this resolution gap, thereby allowing for the study of small-scale cold gas structures evolved in the full $\Lambda$CDM context (see e.g. Fig. 1 of \citealt{pillepich2023}, and \citealt{weinberger2023,butsky2024} (\citealt{hummels2023}) for on the fly (postprocessing) sub-grid approaches to model unresolved cold gas). In particular, the TNG50 simulation has been shown to resolve cold clouds in the gaseous halos of massive ellipticals \citep{nelson2020} and of Milky Way-like galaxies \citep{ramesh2023b,lucchini2024}. However, the simulation resolution limit dictates that only clouds above a mass threshold of $\gtrsim 10^6$\,M$_\odot$ can be studied. Given that this is close to the upper limit of known masses of HVCs \citep{thom2008}, and clouds down to masses of $\lesssim$\,$10^5$\,M$_\odot$ are routinely observed in the Milky Way sky \citep{wakker2001,adams2013}, higher levels of numerical resolution are certainly required to simulate a complete set of analogs of observed CGM clouds. 

Cosmological hydrodynamical simulations that specifically aim to increase numerical resolution in the CGM offer a possible path forwards \citep[e.g.][]{suresh2019,hummels2019,rey2024}. In this work, we use Project GIBLE \citep{ramesh2024a}, a suite of such `CGM-refinement' simulations of ($z$\,$=$\,$0$) Milky Way-like galaxies. The runs we utilize in this work achieve a mass resolution of $\sim$\,$10^3$\,M$_\odot$ in the CGM, i.e. roughly 100 times better than TNG50, allowing us to study smaller structures, while also better resolving clouds in the high-mass end. As we elaborate below, the GIBLE runs also contain a substantial number of Monte Carlo tracer particles, which combined with a snapshot cadence as good as $\sim$\,$30$\,Myr close to $z=0$, makes it possible to quantify the growth and origin of cold clouds to an unprecedented level compared to earlier efforts with cosmological simulations.

The rest of the paper is organised as follows: in Section~\ref{sec:methods}, we provide an overview of the simulation suite, the cloud identification algorithm, Monte Carlo tracers, and the use of these tracers to link clouds between snapshots. Results are presented and discussed in Section~\ref{sec:results}, and summarised in Section~\ref{sec:summary}.

% ------------------------------------------------------------------------------

\section{Methods}\label{sec:methods}

\subsection{Sample and Galaxy Formation Model}

In this paper we use Project GIBLE \citep{ramesh2024a}, a suite of cosmological magneto-hydrodynamical zoom-in simulations of Milky Way-like galaxies with added preferential resolution refinement in the CGM, defined as the region bounded between 0.15\,R$_{\rm{200c}}$ and R$_{\rm{200c}}$ (virial radius). Our sample consists of eight halos, all drawn from the TNG50 Milky Way sample described in detail in \cite{ramesh2023a} \citep[see also][]{pillepich2023}. We achieve CGM gas mass resolutions of $\sim$\,$10^3$, $10^4$ and $10^5$ M$_\odot$ in each of these halos, with the galaxy, i.e. the region within the inner boundary of the CGM (0.15\,R$_{\rm{200c}}$), always maintained at $\sim$\,$8.5 \times 10^5$\,M$_\odot$ (TNG100 resolution). Throughout the paper, we refer to these three resolution levels as `RF512', `RF64' and `R8', respectively. All our main results are derived from the RF512 runs, while the other two are used for resolution convergence tests. In all cases, the simulation output is saved at 30 distinct snapshots: five at relatively-high redshift ($z$\,$\sim$\,[0.50, 0.67, 1.00, 1.50, 2.00]), fifteen between $z$\,$\sim$\,$0.02$ and $z$\,$\sim$\,$0.15$, i.e. with an average snapshot spacing of $\sim$\,$100$\,Myr, and ten between $z$\,$\sim$\,$0.02$ and $z$\,$=$\,$0.0$, i.e. with an even finer time cadence of $\sim$\,$30$\,Myr. The latter ten snapshots are particularly useful to assess short time-scale phenomena, while the next fifteen make it possible to explore the question of cloud growth and evolution over a significant period of time ($\sim$\,$2$\,Gyr). We note that this is typically not possible with cloud crushing setups, which are usually evolved for a few hundred Myr at best, since simulating further is not realistic owing to their idealised nature.

Run with the \textsc{arepo} code \citep{springel2010}, Project GIBLE uses the IllustrisTNG model \citep{weinberger2017,pillepich2018} to account for the physical processes that regulate galaxy formation and evolution. This includes the physics of metal line cooling, star formation and subsequent stellar evolution and enrichment, supermassive black hole (SMBH) formation, and feedback from stars (supernovae) and SMBHs. Included in the TNG model is also ideal magnetohydrodynamics \citep{pakmor2011, pakmor2014} -- a uniform primordial field of $10^{-14}$ comoving Gauss seeded in the initial conditions ($z \sim 127$) is self-consistently amplified through a combination of structure formation, small scale dynamos and feedback processes \citep{pakmor2020,aramburo2021}. Predictions by the TNG model suggest a typical field strength of $1-10 \,\mu\rm{G}$ within galaxies at $z \sim 0$ \citep{marinacci2018, nelson2018, ramesh2023c, ramesh2023a}, consistent with recent observational inferences \citep[e.g][]{mao2017,prochaska2019,heesen2023,boeckmann2023}. 

\subsection{Cloud and Interface Identification Algorithm}

Natural neighbor cells of the unstructured Voronoi tessellation of space provide an ideal spatial definition for clouds. Following \cite{nelson2020} and \cite{ramesh2023b}, we define and identify clouds as spatially contiguous sets of cold ($T \leq 10^{4.5}$\,K) Voronoi cells which contain at least 10 member resolution elements. Clouds composed of fewer than 10 cells are considered unresolved, and are also referred to as cold seeds in subsequent sections of the paper. Cells that are gravitationally bound to satellite galaxies, as identified by the substructure identification algorithm \textsc{subfind} \citep{springel2001} are excised.

The cloud-background interface is defined as the layer of gas cells immediately surrounding the cloud, and is determined using the connectivity of the Voronoi tessellation. It is thus the first cocoon of gas cells around the cloud. Our cloud definition and identification algorithm allows for clouds and interfaces of arbitrary sizes and shapes (see e.g. Fig.~1 of \citealt{ramesh2023b}).

\subsection{Monte Carlo Tracers and Cloud Merger Trees}\label{ssec:merger_tree}

Following an identification of clouds at discrete snapshots, we link clouds with their progenitors and/or descendants with the use of Monte Carlo tracers \citep{genel2013,nelson2015}. In short, these tracers `reside' within resolution elements, and are exchanged between neighbouring cells and/or particles with a probability that scales with the mass flux. As Voronoi cells refine and/or de-refine \citep{springel2010}, tracers are also re-distributed based on the relevant volume ratio. This approach has been shown to be more accurate than velocity field tracers at tracing the underlying mass flow, particularly in regions of strong shocks and/or turbulence \citep{genel2013}.

At the start of a given RF$x$ simulation, we inject $4x$ tracers in every gas cell, resulting in an average of $4$ tracers in every CGM gas cell\footnote{This is $\sim$\,$4$ times more tracers than in the TNG simulations.} following the start of the CGM refinement\footnote{In our simulations, the onset of the refinement takes place once the black hole is seeded ($3$\,$\lesssim$\,$z$\,$\lesssim$\,$5$; \citealt{ramesh2024a}).}. Clouds above the limit of ten member cells thus typically have $\gtrsim$\,40 tracers associated with them. For a given cloud in snapshot $n$, its progenitors in snapshot $n-1$ are defined as clouds that contain at least $1$ of its tracers, and the progenitor with the largest number of tracers is termed the main progenitor. A similar procedure is followed to identify descendants in snapshot $n+1$, and an analogous definition for the main descendant. In effect, we construct the merger trees of clouds (see also \citealt{lucchini2024}), similar to galaxy and subhalo merger tree algorithms \citep[e.g][]{springel2005,behroozi2013,rg2016}. A key difference is that interaction events between clouds may have multiple descendants (during a fragmentation), while this is typically rare for galaxies (see also \citealt{bahe2019}). Note that while cold objects composed of $<$\,10 cells (i.e. unresolved clouds) are included in the tree, mergers and/or fragmentations with them are not counted while computing the associated event rate.

\subsection{Cloud Origins and Mass Flux Rates}

In addition to linking clouds across snapshots and creating their merger trees, we also use tracers to assign one of three origin channels to clouds. Starting from a cloud at a given snapshot (always $z$\,$=$\,$0$ in this work), we follow the main progenitor backwards to a snapshot $n$ such that none of its tracers were present in any CGM cold gas cell in snapshot $n-1$. If the progenitor (in snapshot $n$) is linked to the galaxy or other clouds within $0.15$\,R$_{\rm{200,c}}$ (in snapshot $n-1$), we label the cloud as cold gas arising from the central galaxy. If not, we cross-match all tracers of the progenitor (in snapshot $n$) with those from snapshot $n-1$. The cloud is considered to arise from satellites if \textit{any}\footnote{This is to ensure that we do not `miss' any small cold gas structures from satellites, since tracers rapidly cycle in and out of clouds as they pass through the host CGM \citep{nelson2020}. However, this only leads to a difference in $\lesssim$\,$1$\,$\%$ of cases.} were part of satellite gas. Note that this could refer to gas stripped due to ram-pressure \citep[e.g.][]{rohr2023}, or to those driven out by feedback within the satellite \citep[e.g.][]{peluso2023}. The $z$\,$=$\,$0$ clouds in this category are thus direct descendants of cold gas structures that were ejected from satellites. On the other hand, if the majority were in the non-cold phase of the CGM, we refer to it as a condensate of the hot halo.

Note that while tracers may arise from the IGM, these never form a majority. Similarly, tracers linked to the warm-hot phase of the galaxy (e.g. hot-outflows) never form a majority, although it is possible that the warm-hot gas in the CGM in snapshot $n-1$ is linked to an outflow from an earlier epoch (snapshot $\leq$\,$n-2$). Since the time cadence of the simulation output is coarse beyond $z$\,$\gtrsim$\,$0.15$ (corresponding to a lookback time of $\sim$\,$2$\,Gyr), we terminate the assignment of origin channels at this $z$, and label all still existing clouds as `long lived'. Lastly, at $z$\,$\sim$\,$0.15$, the progenitors of a small fraction ($\lesssim$\,$3$\,$\%$) of $z$\,$=$\,$0$ clouds are below the limit of ten member cells to be defined as a cloud. These are not assigned any of the above flags, and are not considered in any of the subsequent analyses.

Finally, the tracers also enable us to compute mass fluxes between clouds and their environments. `New' tracers in a cloud in snapshot $n$, that were not present in any cold gas cells in snapshot $n-1$, are assumed to trace a mass flux into the cloud from the warm-hot phase. An analogous definition is followed to compute the mass out-flux from a cloud to the surrounding gas. Given the limited time cadence between snapshots, this mass flux need not only represent the cycling of material between the cloud and its interface, but also the flow of material into/from larger scales.

\begin{figure*}[ht!]
    \centering
    \includegraphics[width=18.2cm]{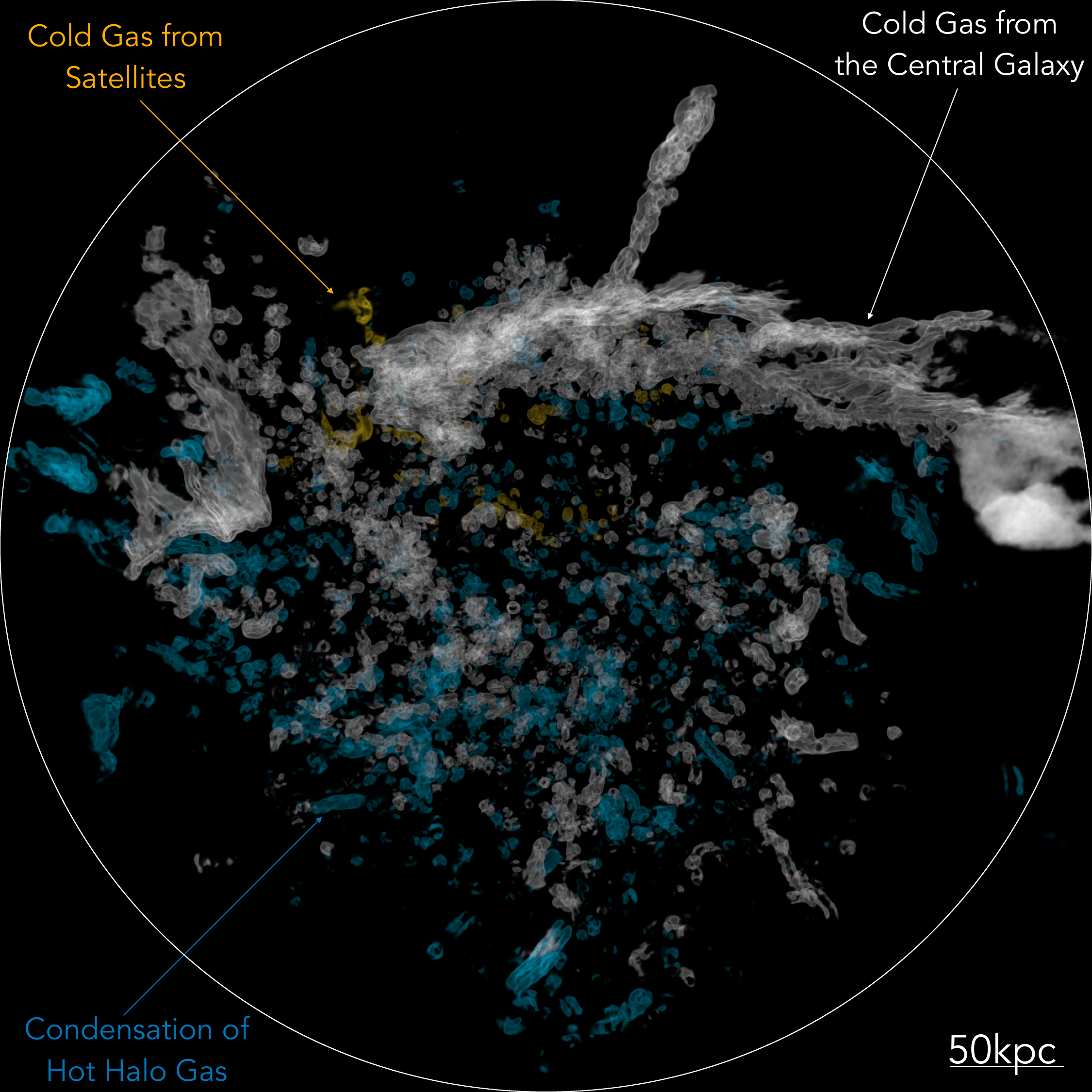}
    \caption{A visual impression of the three origin sources of $z$\,$=$\,$0$ cold CGM gas clouds, with the virial radius of this GIBLE halo (S105; see Table~2 of \citealt{ramesh2024a} for more details) marked using the white circle. Of the many thousand clouds in this CGM, most came to existence in the recent past ($\lesssim$\,$2$\,Gyr). A majority of these can be traced back to cold gas that was previously present in the central galaxy (white). On average across the sample of eight halos, these form $\sim$\,40-60\,$\%$ of the $z$\,$=$\,$0$ population. A small fraction of clouds ($\sim$\,3-5\,$\%$) arise from satellites as they spiral in towards the centre of the halo (orange). Colored in blue, most of the remainder of the clouds are formed through the in-situ condensation of hot halo gas. These account for $\sim$\,15-50\,$\%$, depending on cloud mass. This simple visualisation suggests that the inner regions are dominated by white clouds, while their blue and orange counterparts are situated farther out in the halo. This radial distribution is typical, as we expand upon in subsequent figures.}
    \label{fig:originVis}
\end{figure*}

\begin{figure}[ht!]
    \centering
    \includegraphics[width=9cm]{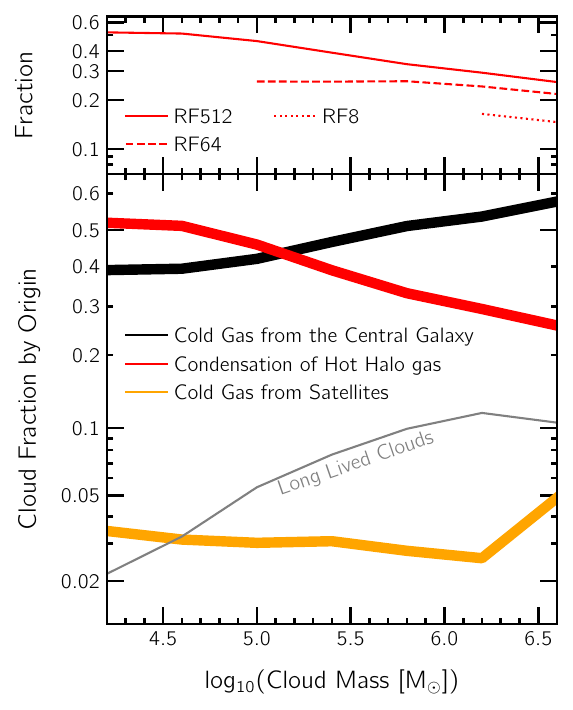}
    \caption{Main panel: The thick solid curves show the fraction of clouds arising from the three origin channels as a function of ($z$\,$=$\,$0$) cloud mass. A large fraction ($\sim$\,40-60\,$\%$) were previously present in the central galaxy, and have been driven out into the CGM in the last $\lesssim$\,$2$\,Gyr (black). Ejection of gas from satellites gives rises to a small portion of the population ($\sim$\,3-5\,$\%$; orange). A majority of the remaining clouds are formed through condensation of the hot phase of the CGM in the last $\sim\,$2\,Gyr. The thin gray curve shows the fraction of clouds that are long lived, i.e. those whose progenitors already exist at least $\sim\,$2\,Gyr ago ($\sim$\,2-10\,$\%$). In the top panel, we assess numerical convergence. The fraction of clouds arising through in-situ condensation of the hot-phase in the RF512 suite (solid curves) is contrasted against the lower resolution RF64 (dashed) and RF8 (dotted) runs. Results between resolution levels approach convergence towards the high mass end, i.e. well above the resolution limit.}
    \label{fig:originFraction}
\end{figure}

\begin{figure*}[ht!]
    \centering
    \includegraphics[width=18cm]{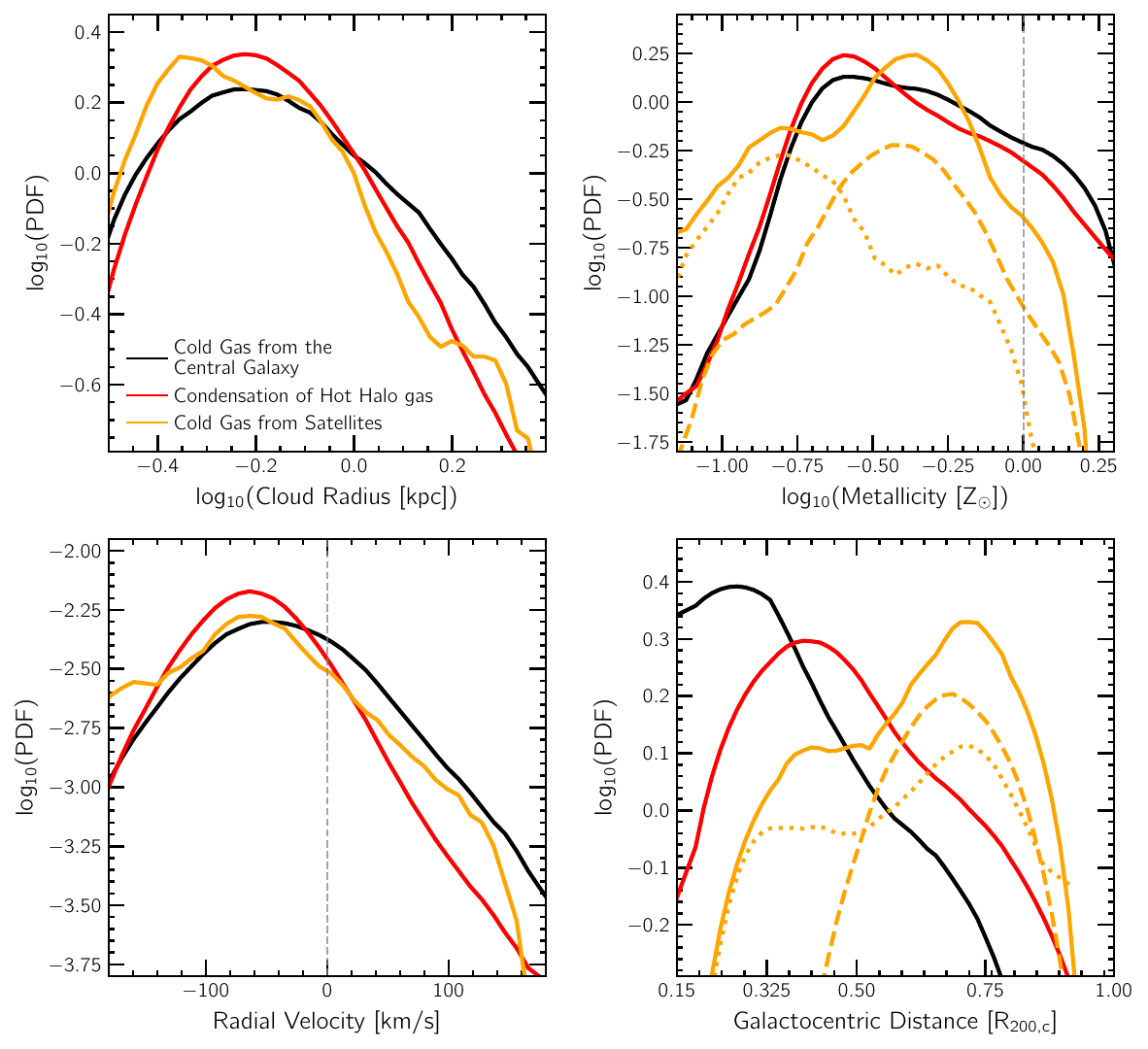}
    \caption{PDFs of properties of $z$\,$=$\,$0$ CGM clouds, split based on their origin channel. Curves corresponding to those arising as from the central galaxy (within the last $\sim$\,$2$\,Gyr) are colored in black, condensates of the hot halo gas in red and those ejected from satellites in orange, as shown by the legend in the top-left panel. Distributions of cloud radius, metallicity, radial velocity and galactocentric distance are shown in the top-left, top-right, lower-left and lower-right panels, respectively. In the right panels, we show individual distributions of the S98 and S105 halos (see Table~2 of \citealt{ramesh2024a} for more details) through dotted and dashed curves, respectively. Distributions are broad for all four properties shown here, with PDFs of clouds ejected from satellites typically different as compared to the other two cases.}
    \label{fig:origin_vs_prop}
\end{figure*}

\begin{figure}[ht!]
    \centering
    \includegraphics[width=9cm]{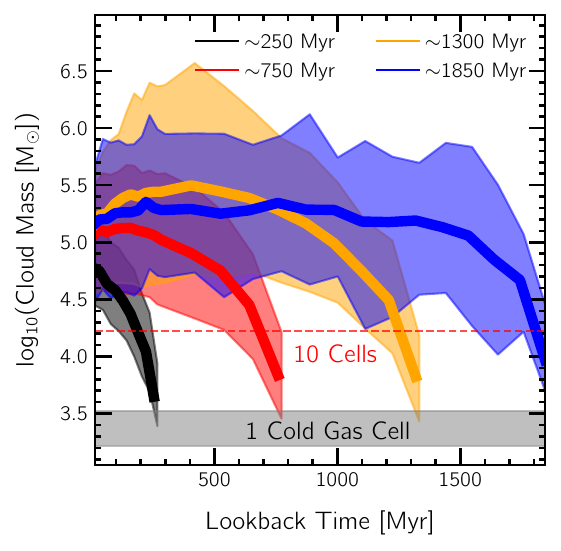}
    \caption{Evolutionary tracks of clouds that condense out of the hot halo. Solid curves correspond to the median while the shaded bands show the 16$^{\rm{th}}$-84$^{\rm{th}}$ percentile regions. Colors correspond to different epochs where the cloud first came into existence. In all cases, the origin of the cloud can be traced back to a `cold seed', typically comprised of only a few gas cells. Following that, clouds accrete matter to grow to a median mass of $\sim$\,$10^{5.2}$\,M$_\odot$ by $z$\,$=$\,$0$. The percentile regions, however, are broad, implying that evolutionary tracks of individual clouds can be diverse.}
    \label{fig:cldEvlCon}
\end{figure}

\begin{figure}[ht!]
    \centering
    \includegraphics[width=9.01cm]{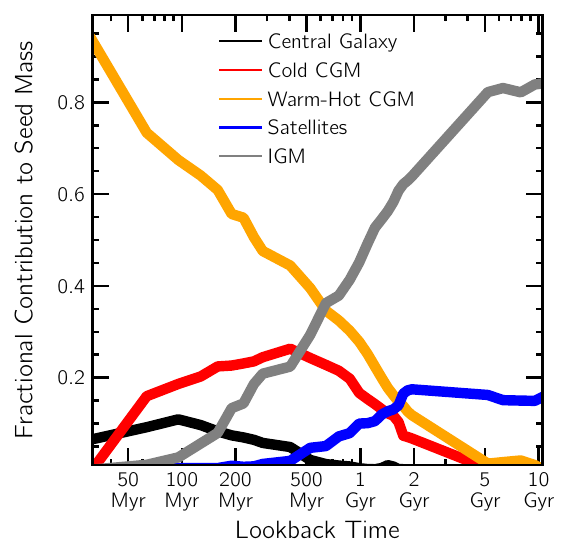}
    \caption{Dissecting the different sources of gas that seed the condensation of cold clouds out of the hot halo. On short time-scales ($\lesssim$\,$500$\,Myr), the dissolution of past clouds into the hot phase, and the subsequent mixing, is the primary source, potentially also stirred by outflows from the central galaxy into the CGM. Transitioning out into previous cosmic epochs as early as $z$\,$\sim$\,$2$ (t$_{\rm{lookback}}$\,$\sim$\,$10.5$\,Gyr), the mass giving rise to these seeds was predominantly smoothly accreted from the IGM into the halo ($\sim$\,$85$\,$\%$), while clumpy accretion through satellites contributes the remaining $\sim$\,$15$\,$\%$.}
    \label{fig:seedOrigin}
\end{figure}

\begin{figure}[ht!]
    \centering
    \includegraphics[width=9cm]{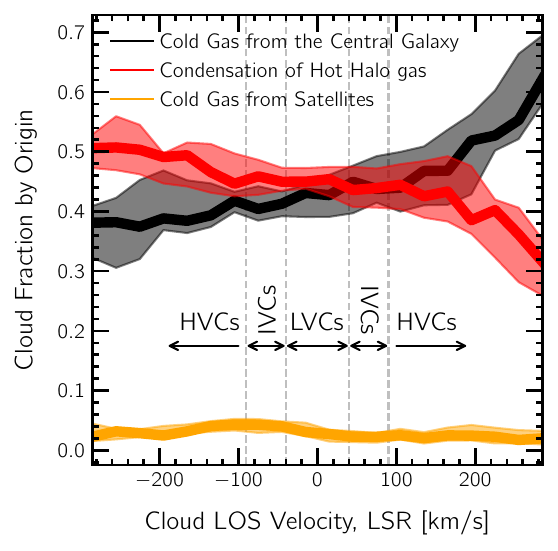}
    \caption{Probabilities of the three origin channels as a function of line-of-sight (LOS) velocities of clouds in the frame of the local standard of rest (LSR). The vertical dashed lines delineate commonly used definitions to classify clouds in the Milky Way halo. Outflowing clouds predominantly trace cold gas arising from the central in the recent past ($\lesssim$\,$2$\,Gyr), while their inflowing counterparts typically condense out of the hot phase. Clouds ejected from satellites are relatively rare ($\lesssim$\,5\,$\%$), but roughly equally likely at all (LOS) velocities.}
    \label{fig:originFraction_vs_vlos}
\end{figure}

% ------------------------------------------------------------------------------

\section{Results}\label{sec:results}

\subsection{The Origin of $z$\,$=$\,$0$ Clouds}\label{ssec:origin}

We begin with a visualisation\footnote{Isosurface volume rendering made with \href{https://github.com/nelson-group/ArepoVTK}{ArepoVTK.}} highlighting the different origin channels defined above. Fig.~\ref{fig:originVis} shows the distribution of clouds in the CGM of a random GIBLE halo at $z=0$. The image stretches $\pm$\,R$_{\rm{200,c}}$ from edge-to-edge, where R$_{\rm{200,c}}$ is the virial radius of the halo, also marked by the white circle. While the projection extends across the same range along the perpendicular direction, clouds within a galactocentric distance of $0.15$\,R$_{\rm{200,c}}$, i.e. inside the inner boundary of the CGM, are not considered in this work, and are not shown here. The clouds are colored by the different origin channels assigned to them: those that arise from the central galaxy are shown in white. Condensates of the hot CGM phase are shown in blue, while those ejected from satellites are colored in orange. Note that the long-lived clouds in this halo are not shown in this image. This simple visualisation suggests that the white and blue channels dominate, while clouds arising through stripping of satellite gas are less abundant.

In Fig.~\ref{fig:originFraction}, we quantify this visual inference by showing the fraction of clouds originating from each channel, as a function of $z$\,$=$\,$0$ cloud mass. The cloud fraction arising from the galaxy is shown in black, condensation of the hot halo in red, and ejected from satellites in orange. To increase statistics, we combine all clouds across all eight GIBLE halos.

As suggested by Fig.~\ref{fig:originVis}, clouds from the central galaxy and hot halo condensation dominate. On average, $\sim$\,$40$\,$\%$ of $\sim$\,$10^{4.2}$\,M$_\odot$ clouds at $z$\,$=$\,$0$ arise as a result of cold gas transition from the into the CGM in the last $\sim$\,$2$\,Gyr. This number increases with cloud mass, to as high as $\sim$\,$60$\,$\%$ for their massive counterparts ($\sim$\,$10^{6.6}$\,M$_\odot$). While $\sim$\,$67$\,$\%$ of Milky Way-like galaxies in TNG50 are in the regime where kinetic kicks by the central super-massive black hole (SMBH) is the major mode of feedback at $z$\,$\lesssim$\,$0.15$ \citep{pillepich2021,ramesh2023a}, \textit{all} eight halos re-simulated in Project GIBLE are in the kinetic mode by $z=0$. In our sample, outflowing clouds from the galaxy into CGM over the last $\sim$\,$2$\,Gyr are primarily driven by SMBH i.e. AGN feedback. Note, however, that some of these clouds may not be ejected due to feedback processes, but rather be dislodged from the extended or outer disk due to rapid rotation and dynamics alone. Furthermore, galactic-scale winds driven by supernovae feedback will also produce cool outflowing gas \citep{fraternali2006,smith2024}, a phenomenon more readily apparent at higher redshifts and/or lower galaxy masses \citep[see also][]{suresh2019}.

At the low-mass end ($\sim$\,$10^{4.2}$\,M$_\odot$), clouds are more likely to arise through hot-phase condensation ($\sim$\,$50$\,$\%$). This origin channel portrays the inverse mass trend as above, with the fraction dropping to $\sim$\,$25$\,$\%$ as one transitions to the high mass end ($\sim$\,$10^{6.6}$\,M$_\odot$). While theory suggest that such precipitation can take place if the cooling time of gas is sufficiently short \citep{mccourt2012,voit2017}, this condition is typically not satisfied (globally) across the halo \citep{nelson2020}. However, strong density perturbations can lead to run away thermal instability and cooling \citep{sharma2012a,choudhury2019,dutta2022}, the exploration of which we return to shortly.

Descendants of cold gas structures ejected from satellites are relatively rare. The fraction of such clouds is relatively flat at $\sim$\,2-3\,$\%$ in the mass range of $\sim$\,10$^{4.2}$-10$^{6.2}$\,M$_\odot$, with a slightly larger fraction of $\sim$\,5\,$\%$ for clouds as massive as $\sim$\,$10^{6.6}$\,M$_\odot$. While past studies have shown that stripping of satellite gas may substantially add to the cold mass budget of the host CGM \citep[e.g.][]{olano2008,yun2019}, only four of the eight halos in our sample have had gaseous satellites in-fall through them in the last $\sim$\,$2$\,Gyr, thus possibly leading to an under-estimate. Indeed, for a larger sample of $132$ Milky Way halos in TNG50, clouds preferentially cluster along the past trajectories of satellites \citep{ramesh2023b}.

In addition, we note that the total mass of these satellites at first infall was $\sim$\,[$10^{10.7}$, $10^{10.8}$, $10^{11.4}$, $10^{11.45}$]\,M$_\odot$. All four first crossed the halo virial radius between $z$\,$\sim$\,$0.15$ and $0.5$, and owing to the coarse time resolution beyond $z$\,$\sim$\,$0.15$, it is likely that these numbers are lower limits. In either case, the latter two of these interactions resulted in major mergers \citep[merger mass ratio greater than $1/4$, defined as the ratio of stellar masses when the secondary has its maximum mass;][]{rg2016}. It is thus possible that while the CGM of these satellites was stripped as they spiralled in to the central galaxy, only a fraction of their cold-phase ISM was \citep[e.g.][see also discussion below]{ghosh2024}. This would reduce the fraction of clouds arising through these interactions.

The thin gray curve shows the fraction of long lived clouds, i.e. those whose progenitors were already assembled $\sim$\,$2$\,Gyr ago. The fraction of such clouds is relatively small ($\sim$\,$2$\,$\%$) at the low mass end ($\sim$\,$10^{4.2}$\,M$_\odot$), and gradually increases to $\sim$\,$10$\,$\%$ for their more massive counterparts ($\sim$\,$10^{6.6}$\,M$_\odot$). Larger clouds are thus more likely to survive over a long time scale. Studies with idealised simulations have reached a similar conclusion, arguing that clouds survive above some critical minimum mass \citep{gronke2022}. The evolution of long-lived clouds is a main focus of our current work, which we return to and explore in Section~\ref{ssec:evl_long_lived}.

In the top panel, we assess numerical convergence. To do so, we compare the fraction of clouds with a condensation origin in the RF512 suite (solid line) with the lower resolution RF64 (dashed) and RF8 (dotted) runs. Note that the latter two curves only begin at $\sim$\,$10^5$ and $10^6$ respectively, since clouds below this threshold are unresolved in these runs. While there is a clear offset between RF512 and RF64 at low masses ($\sim$\,$10^5$\,M$_\odot$), they approach convergence towards the high mass end, i.e. for massive clouds well above the resolution limit. The RF8 run, i.e. TNG50-1 resolution, is far from convergence at these mass scales, but approaches convergence for higher mass scales, although such clouds are rare in the sample \citep{ramesh2024a}.

Before proceeding further, we contrast our findings against a recent study that likewise aimed to quantify the origin of cold circumgalactic gas \citep{decataldo2023}. This study is based on a single Milky Way like halo `Eris2k' at $z$\,$\sim$\,$0.3$, simulated with the SPH code \textsc{gasoline} \citep{wadsley2004} at an average baryonic mass resolution of $\sim$\,$2$\,$\times$\,$10^4$\,M$_\odot$, i.e. a factor of $\sim$\,$10$ coarser than our RF512 runs. Their results suggest that condensation out of the hot phase is the primary source of cold gas ($\sim$\,$80$\,$\%$), with contributions by satellites ($\sim$\,$2$\,$\%$) and outflows from the galaxy ($\sim$\,$15$\,$\%$) being less prominent. Apart from the differences in redshift and resolution between Eris2k and GIBLE, we note that the underlying numerical methods and feedback recipes are also different. We speculate that a combination of these factors is the leading cause for differing results. Moreover, while we show results for clouds stacked across all halos in Fig.~\ref{fig:originFraction}, we mention that the fractions of the two dominant channels typically vary by $\sim$\,$\pm$\,$15$\,$\%$ across the set of halos. A larger sample size with the Eris2k model may change their average result.

In Fig.~\ref{fig:origin_vs_prop}, we explore physical properties of $z$\,$=$\,$0$ clouds, split based on the three origin channels discussed above. Similar to Fig.~\ref{fig:originFraction}, properties of clouds arising from satellites are shown in black, condensation of the hot halo in red, and ejected from satellites in orange. Clouds across the sample are again stacked.

In the top-left panel, we show distributions of cloud sizes, measured as the radius of the volume equivalent sphere. While both the black and red curves peak around $\sim$\,$650$\,pc, the former is more broad, extending to sizes as large as $\gtrsim$\,$2$\,kpc. This is a result of clouds in the latter category being more spherical, as opposed to elongated, with smaller major-to-minor axis ratios on average (not shown explicitly, although visible in Fig.~\ref{fig:originVis}). Clouds ejected from satellites are typically smaller, peaking at $\sim$\,$450$\,pc. We interpret that this is due to densities being larger in the ISM of satellites than in the cold-phase CGM, leading to smaller clouds for a given mass. The distribution, however, extends to as large as $\gtrsim$\,$2$\,kpc, similar to the red curve, suggesting that this may be a signature of cold CGM gas of satellites being stripped, as we explore below.

The top-right panel shows PDFs of cloud metallicity. The red and black curves are largely similar, peaking at $\sim$\,$25$\,$\%$\,Z$_\odot$. The distributions are, however, broad, with clouds as enriched as $\sim$\,$2$\,Z$_\odot$, and as pristine in their metal content as $\lesssim$\,$0.1$\,Z$_\odot$. The likeness between the two curves suggests a substantial level of mixing between halo gas and metal enriched outflows driven outwards from the galaxy by feedback processes \citep[][see also Appendix~\ref{app:accrn}]{fielding2017,nelson2019}. A similar variance of metallicity is observed for high velocity clouds (HVCs) in the Milky Way halo, with values ranging between $\lesssim$\,$0.3$\,Z$_\odot$ \citep{richter2001,collins2003} to $\gtrsim$\,$1.5$\,Z$_\odot$ \citep{zech2008,yao2011}.

The orange curve, however, portrays a different behaviour: it is bimodal, with a primary peak at $\sim$\,$0.45$\,Z$_\odot$, and a smaller secondary peak at a lower value of $\sim$\,$0.15$\,Z$_\odot$. This is a result of superposing the distributions from the different halos with their distinct satellite populations. To dig deeper, we also show the individual distributions from two particular halos: the dashed curve corresponds to the halo with a $\sim$\,$10^{11.40}$\,M$_\odot$ satellite discussed above, while the dotted line shows the halo with the $\sim$\,$10^{11.45}$\,M$_\odot$ satellite, both vertically offset downwards for visibility. Clouds from these two halos dominate the orange curve, so considering these alone is sufficient to explain its structure. The dashed line peaks at higher metallicity, suggesting that these clouds are signatures of ISM gas being stripped. On the contrary, the dotted curve peaks at lower metallicity, and we interpret that these clouds may instead primarily arise from the cold CGM of this satellite. At the same time, a small feature at higher $Z$ hints that a portion of the ISM is also stripped, as we discuss more below. The low-$Z$ tail of clouds arising from satellites is also more prominent compared to the black and red curves, suggesting that much of the CGM gas of these satellites is metal poor in comparison to the CGM of the central galaxy \citep{pasquali2012}.

In the lower-left panel, we show distributions of radial velocities. The black curve peaks at $\sim$\,$-40$\,km/s, i.e. most clouds that moved outwards from the galaxy in the last $\sim$\,$2$\,Gyr are now inflowing, tracing the second phase of the halo fountain cycle \citep{peroux2020}. The distribution, however, extends to large positive radial velocities, with clouds moving from the galaxy into the CGM at speeds as large as $\sim$\,$200$\,km/s, corresponding to the first phase of a halo-scale fountain flow \citep{oppenheimer2008}. 

The red curve peaks at a comparatively larger negative radial velocity of $\sim$\,$-60$\,km/s, i.e. a larger fraction of clouds are inflowing as compared to the black curve, as they precipitate out of the hot halo and begin infalling under the influence of gravity \citep{voit2015a}. Although outflowing clouds in this category are less common as compared to the black curve, a small fraction of clouds are moving away from the galaxy, with outward velocities as large as $\sim$\,$200$\,km/s. These are clouds that likely precipitated out of warm-hot outflows from the galaxy, or from galactic cold gas that dissolved in the surrounding wind and thereafter condensed \citep[e.g.][see also Fig.~\ref{fig:seedOrigin}]{zhang2017}. Clouds stripped from satellites are also preferentially inflowing, with the distribution peaking at $\sim$\,$-60$\,km/s. Some clouds are, however, outflowing, either as a result of their rotational vectors aligning with the direction of ram pressure \citep{sparre2024}, or possibly due to feedback processes within the satellite driving gas along the stripped tail \citep{peluso2023}.

Lastly, in the lower-right panel, we focus on the distributions of galactocentric distance. The black curve is relatively flat between the inner radius of the CGM ($0.15$\,R$_{\rm{200,c}}$) and $\sim$\,$0.3$\,R$_{\rm{200,c}}$, and begins to decline towards larger distances, i.e. these clouds are centrally concentrated. The red curve peaks at $\sim$\,$0.4$\,R$_{\rm{200,c}}$: these clouds dominate the inner halo. However, the distribution is broad, with clouds as far as $\sim$\,$0.9$\,R$_{\rm{200,c}}$, but also close to the inner boundary of the CGM. Similar to the top-right panel, clouds ejected from satellites show a bimodal behaviour, with peaks at $\sim$\,$0.4$ and $\sim$\,$0.8$\,R$_{\rm{200,c}}$, once again as a result of superposing distributions across halos. The dashed curve (the $\sim$\,$10^{11.4}$\,M$_\odot$ satellite) peaks at relatively large galactocentric distances, suggesting that this satellite was stripped soon after infall. The dotted curve (the $\sim$\,$10^{11.45}$\,M$_\odot$ satellite) also shows a primary peak at large galactocentric distances, and this corresponds to the low-metallicty CGM gas. A second less prominent peak in the inner region of the halo is seen, signifying the onset of stripping of ISM gas as ambient densities grow larger \citep[][]{ayromlou2019}. Furthermore, these two satellites with roughly the same total mass at infall seem to loose their gas in varying ways, suggesting that orbital diversity and large-scale environment play important roles in such hydrodynamic interactions \citep[e.g.][]{roediger2007,sales2015}.

\vspace{-0.5cm}

\textit{\subsubsection{Precipitation of the Hot Phase}}

As discussed above, cold clouds of gas may be able to condense out of the hot phase of the CGM \citep[e.g.][]{fraternali2006}. Indeed, this is the origin channel for $\sim$\,25-50\,$\%$ of clouds in our sample (Fig.~\ref{fig:originFraction}). In this sub-section, we take a deeper look into their growth.

We begin with evolutionary tracks of mass growth in Fig.~\ref{fig:cldEvlCon}. We group clouds into four bins: those that first came into existence as cold CGM gas cells\footnote{At this stage, note they may not yet be classified as clouds, as they could be below the threshold of 10 member cells.} $\sim$\,250, 750, 1300 and 1850\,Myr ago, shown in black, red, orange and blue, respectively. These numbers are thus also upper limits on the formation times of their progenitors. Solid curves show the median, while the shaded bands correspond to the 16$^{\rm{th}}$-84$^{\rm{th}}$ percentile regions. As before, we stack clouds across all halos. The gray band shows the mass range within which gas cells are maintained by \textsc{arepo} through refinement and de-refinement, i.e. corresponding to the mass of one gas cell, while the red dashed line shows ten times the average mass resolution.

Going backwards in time starting from $z$\,$=$\,$0$, all evolutionary tracks lead back to a small cold gas object, typically only comprised of a few cells. The fact that it is not always \textit{one} cell at the start is likely due to limited time resolution between snapshots.\footnote{The black curve is derived from snapshots with superior time resolution. Comparing it to the others, we see that the mass of the initial object is indeed lower. Appendix~\ref{app:time_resl} further discusses tests related to limited time resolution.} In all cases, starting from this cold over-density, curves grow steadily to reach a median mass of $\sim$\,$10^{5.2}$\,M$_\odot$ by $z$\,$=$\,$0$. The percentile regions are however broad, implying that evolutionary tracks may be diverse. Furthermore, while the median curves are smooth and roughly monotonous, the growth of individual objects is typically not (see upper panels of Fig.~\ref{fig:cldMass_evl} for analogous growth tracks).

Precipitation of clouds out of the hot phase thus begins with the formation of a `cold seed'. Such an idea was discussed earlier by \cite{nelson2020}, where it was suggested that thermal instability and runaway cooling could thereafter lead to accretion of cold material \citep[e.g.][]{field1965,sharma2012b,dutta2022}, ultimately resulting in the growth of the seed into a bigger cloud. However, it is presently unclear as to how such a central cold over-density is formed in the first place.

We aim to address this question in Fig.~\ref{fig:seedOrigin}. We focus on seeds that first appeared at $z$\,$=$\,$0$, and note that the result is similar for seeds that appeared at higher $z$. To boost statistics, we stack tracers across all seeds, and also across all halos. The y-axis shows the fractional contribution to the mass of these seeds as a function of lookback time, all the way back to $z$\,$\sim$\,$2$ (t$_{\rm{lookback}}$\,$\sim$\,$10.5$\,Gyr). We consider five distinct sources: (a) gas from the central galaxy (all phases of gas included) shown in black, (b) the cold CGM, i.e. clouds that existed in the past but subsequently dissolved into the hot phase, in red, (c) the warm-hot CGM in orange, (d) gas from satellites i.e. other galaxies in general\footnote{Although we label this category as satellites, these galaxies could be centrals of their own distinct halos at high-$z$.} (all phases included) in blue, and (e) the IGM\footnote{We define the IGM to be gas outside R$_{\rm{200,c}}$ of the main galaxy, that is not gravitationally bound to any halo.} in gray.

At a lookback time of $\sim$\,$30$\,Myr, i.e. just before the seed was formed, $\sim$\,$95$\,$\%$ of the mass was in the warm-hot phase of the CGM. The remainder ($\sim$\,$5$\,$\%$) arose from the central galaxy, which by definition corresponds to warm-hot gas being ejected outwards. While the contribution of other cold CGM clouds is $0$\,$\%$ at this first snapshot, it rises to as large $\sim$\,$25$\,$\%$ at earlier times. The dissolution of clouds into the background halo, and the eventual mixing between phases, thus plays an important role in seeding the next generation of clouds \citep[see also][]{nelson2020}. In addition, the mass contribution due to outflows\footnote{At t$_{\rm{lookback}}$\,$>$\,$30$\,Myr, i.e. at least two snapshots prior, cold gas outflows can contribute, although this component is always $\lesssim$\,$10$\,$\%$.} from the galaxy rises to as high as $\sim$\,$10$\,$\%$. Turbulence driven by the interaction of these outflows with the CGM gas may further contribute to precipitation \citep{mohapatra2019}.

Zooming out to previous cosmic epochs as early as $z$\,$\sim$\,$2$, the mass contribution of IGM gas rises to $\sim$\,$85$\,$\%$. A similar increase in the contribution of gas from other galaxies is also seen, to $\sim$\,$15$\,$\%$ at t$_{\rm{lookback}}$\,$\gtrsim$\,$2$\,Gyr. Almost all of the gas seeding condensation at $z$\,$=$\,$0$ (and in general low-$z$) is thus extra-galactic at $z$\,$\sim$\,$2$, after which it accretes, either smoothly or in a clumpy fashion \citep{keres2005,nelson2013}. At least some of this gas ultimately reaches the galaxy (Appendix~\ref{app:accrn}), and is recycled back into the CGM, possibly multiple times \citep{alcazar2017,suresh2019}. This halo fountain, as well as the motion of satellite galaxies through the CGM, likely produces additional density perturbations in the halo \citep{gauthier2013,osullivan2018}, giving rise to clouds that thereafter dissolve in the hot phase, eventually leading to a new population of clouds, and so on. Overall, the short time-scale ($\lesssim$\,$500$\,Myr) dissolution of pre-existing cold structures into the hot phase primarily drives the formation of cold seeds. Simultaneously, the source of their baryonic mass is ultimately of cosmological origin (see also Appendix~\ref{app:accrn}).

\textit{\subsubsection{Predicting the Origin of Milky Way Clouds}}

We conclude the origin-segment of this paper with a simple prediction for the origin of high velocity clouds (HVCs), and in general cold clouds, in the Milky Way halo. Following \cite{ramesh2023b}, we place a hypothetical observer at a random point along the galactic plane, at a galactocentric distance of $8.34$~kpc. Further, the observer is considered to be moving along a perfectly circular orbit around the galactic centre, at a velocity of $240$~km/s. This observer is consistent with the solar location and motion in the Milky Way \citep{reid2014}. We repeat this randomization $100$ times for each halo, resulting in a total of $800$ random realisations for the eight GIBLE halos stacked together. 

In Fig.~\ref{fig:originFraction_vs_vlos}, we show the fraction of cloud origin as a function of the line-of-sight (LOS) velocity to this hypothetical observer. Clouds arising from the galaxy are shown in black, condensation of the hot halo in red, and ejected or stripped from satellites in orange. Solid lines show the median, and the shaded bands the 16$^{\rm{th}}$-84$^{\rm{th}}$ percentile regions across the random realisations. As the ultra-refined region in our simulation is bounded by $0.15$\,R$_{\rm{200,c}}$ ($\sim$\,30-40\,kpc), only clouds beyond this distance are considered. Known distances to HVCs are typically $\sim$\,4-15\,kpc \citep{wakker2008,peek2016}, although these are estimated using absorption-line spectra along the line-of-sight of background stars, and thus the lack of known clouds at larger distance estimates is likely a direct result of the dearth of distant halo stars, i.e. an observational bias.

The different vertical dashed lines and labels correspond to commonly used definitions to classify clouds based on their kinematics: LVCs (low velocity clouds) are those with (absolute) line of sight velocities in the range $\lesssim$\,$40$\,km/s, IVCs (intermediate velocity clouds) with 40-90\,km/s, and HVCs with $\gtrsim$\,$90$\,km/s \citep[e.g.][]{wakker2001}. 

Fig.~\ref{fig:originFraction_vs_vlos} suggests that LVCs and IVCs are roughly equally likely to arise as a result of cold gas moving outwards from the galaxy, as well as through condensation of the hot phase. At larger velocities, the two median curves begin to separate out. Specifically, at line-of-sight velocities $\lesssim$\,$-100$\,km/s, the red curve overtakes the black, i.e. inflowing HVCs predominantly trace clouds that condense out of the hot halo. On the contrary, at velocities $\gtrsim$\,$100$\,km/s, the inverse trend is observed, i.e. outflowing HVCs are more likely descendants of cold gas structures that were ejected from the galaxy. Note, however, that the two percentile regions overlap out to velocities as large as $\sim$\,$\pm$\,$200$\,km/s, particularly in the regime of positive velocities. Unambiguous identification of the origin of clouds in the Milky Way halo is thus not possible with kinematics alone.

Clouds ejected from satellites are relatively rare ($\lesssim$\,$5$\,$\%$), as previously discussed, but more or less equally likely at all line-of-sight velocities. Note, however, that these fractions are for clouds stacked across all masses and distances, and numbers may change if the selection of clouds is restricted based on a given metric. For instance, the fraction of clouds arising from the cold gas supply of the galaxy is a bit larger in the inner halo, as discussed earlier in the lower-right panel of Fig.~\ref{fig:origin_vs_prop}. 

Considering additional physical properties beyond kinematics could help break degeneracies and more clearly identify cloud origins. Metallicity and metal abundance ratios are one example. Mock absorption spectra through clouds and their interfaces, as a function of origin, will establish mappings between observable and unobservable cloud properties \citep[e.g.][]{eisert2024,weng2024} in future work (\textcolor{blue}{Guo et al. in prep}).

\vspace{0.5cm}

\begin{figure*}[ht!]
    \centering
    \includegraphics[width=18.2cm]{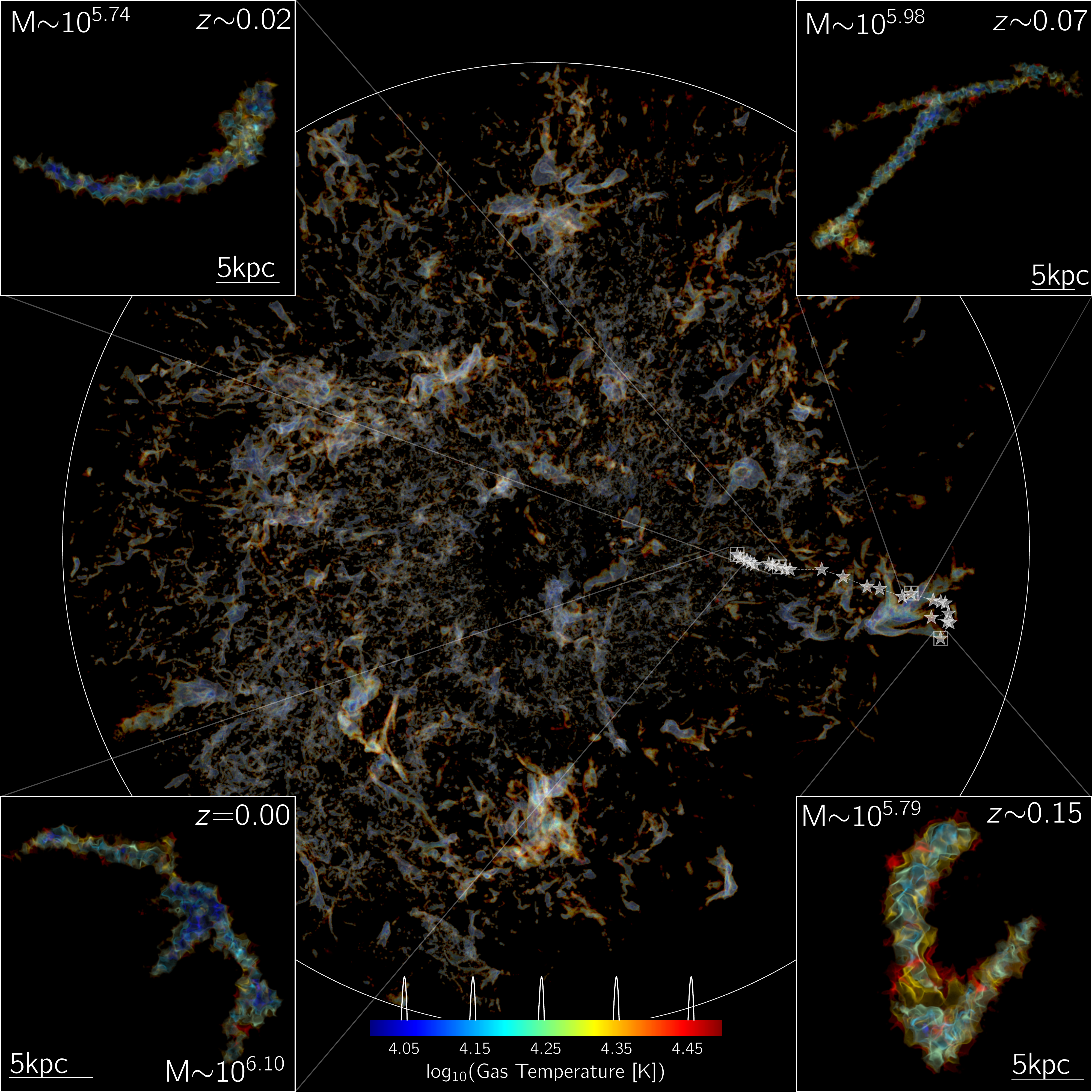}
    \caption{A visualisation of cold clouds in a GIBLE halo (S98; see Table~2 of \citealt{ramesh2024a} for more details) at $z$\,$=$\,$0$ using a ray-traced volume rendering sampled by five narrow Gaussian transfer functions in temperature space, as shown by the color bar (background image). The white circle denotes the outer boundary of the CGM refinement region, i.e. the virial radius of the halo ($\sim$\,$270$\,kpc). In the foreground, we show the (past) trajectory of a $\sim$\,$10^{6.1}$\,M$_\odot$ cloud identified at $z$\,$=$\,$0$, with the various translucent stars highlighting the positions of the main progenitor of this cloud at $25$ distinct snapshots back to $z$\,$\sim$\,$0.15$. The insets at the corners of the image show a volume rendering of (the main progenitor of) this cloud at four distinct redshifts, going back in time in clockwise direction starting from the bottom left, with the cloud oriented such that its mean velocity vector points to the right. In the $\sim$\,$2$\,Gyr between $z$\,$\sim$\,$0.15$ and 0, this cloud clearly evolves in terms of shape, size (characterized by the varying scale bars in the insets) and mass (shown by the different labels; in units of M$_\odot$). Also visible in the top-right inset is an instance of an ongoing merger between clouds, a phenomenon that will be explored and quantified later in this paper.}
    \label{fig:evlVis}
\end{figure*}

\begin{figure*}[ht!]
    \centering
    \includegraphics[width=18cm]{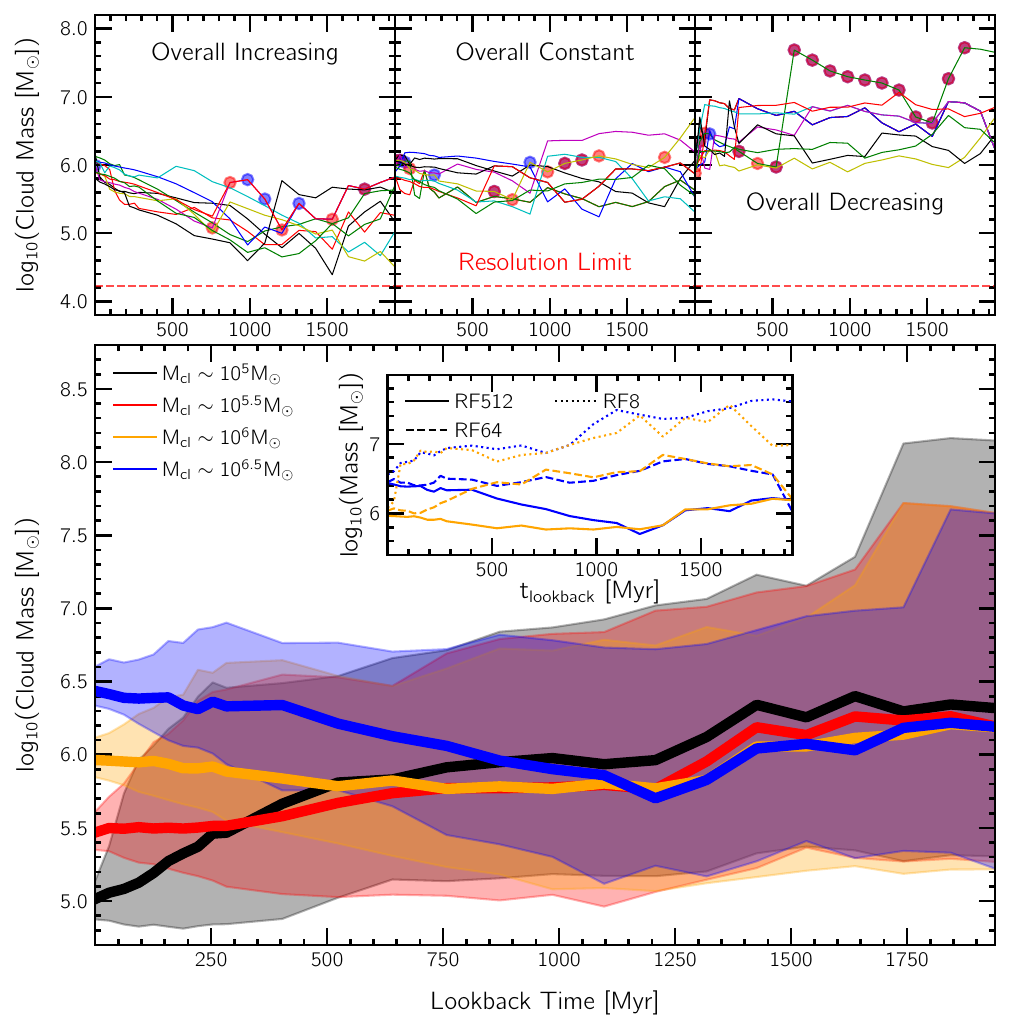}
    \caption{Main panel: evolution of masses of $z$\,$=$\,$0$ clouds whose main progenitors already existed $\sim$\,$2$\,Gyr ago ($z$\,$\sim$\,$0.15$). Solid curves correspond to medians, shaded bands to 16$^{\rm{th}}$-84$^{\rm{th}}$ percentile regions, and the colors to four different mass bins. On average, progenitors of clouds from all four bins were massive ($\sim$\,$10^{6.25}$\,M$_\odot$) $2$\,Gyr ago, albeit with relatively broad percentile regions in all cases. In the top panel, we show individual tracks for a subset of clouds split into three categories: those whose masses have overall increased (left), remained overall constant (centre) and overall decreased (right) in this time frame. For exactly one curve in each case, we show all instances of mergers and fragmentations as identified in our cloud merger trees: the former are shown in blue dots, the latter in red, and overlapping dots of both colors correspond to instances where both these events take place between neighbouring snapshots. As expected, sudden jumps in curves are associated with these events. Finally, in the inset of the main panel, we assess numerical convergence by contrasting results of RF512 (solid) with the lower resolution RF64 (dashed) and RF8 (dotted) runs for the case of ($z$\,$=$\,$0$) $\sim$\,$10^{6}$ and $10^{6.5}$\,M$_\odot$ clouds. While the three runs are qualitatively similar, curves converge at a higher mass for the lower resolution runs (see text for discussion). Note that the axes labels of the inset and the top panels are identical to that of the main panel.}
    \label{fig:cldMass_evl}
\end{figure*}

\begin{figure*}[ht!]
    \centering
    \includegraphics[width=18cm]{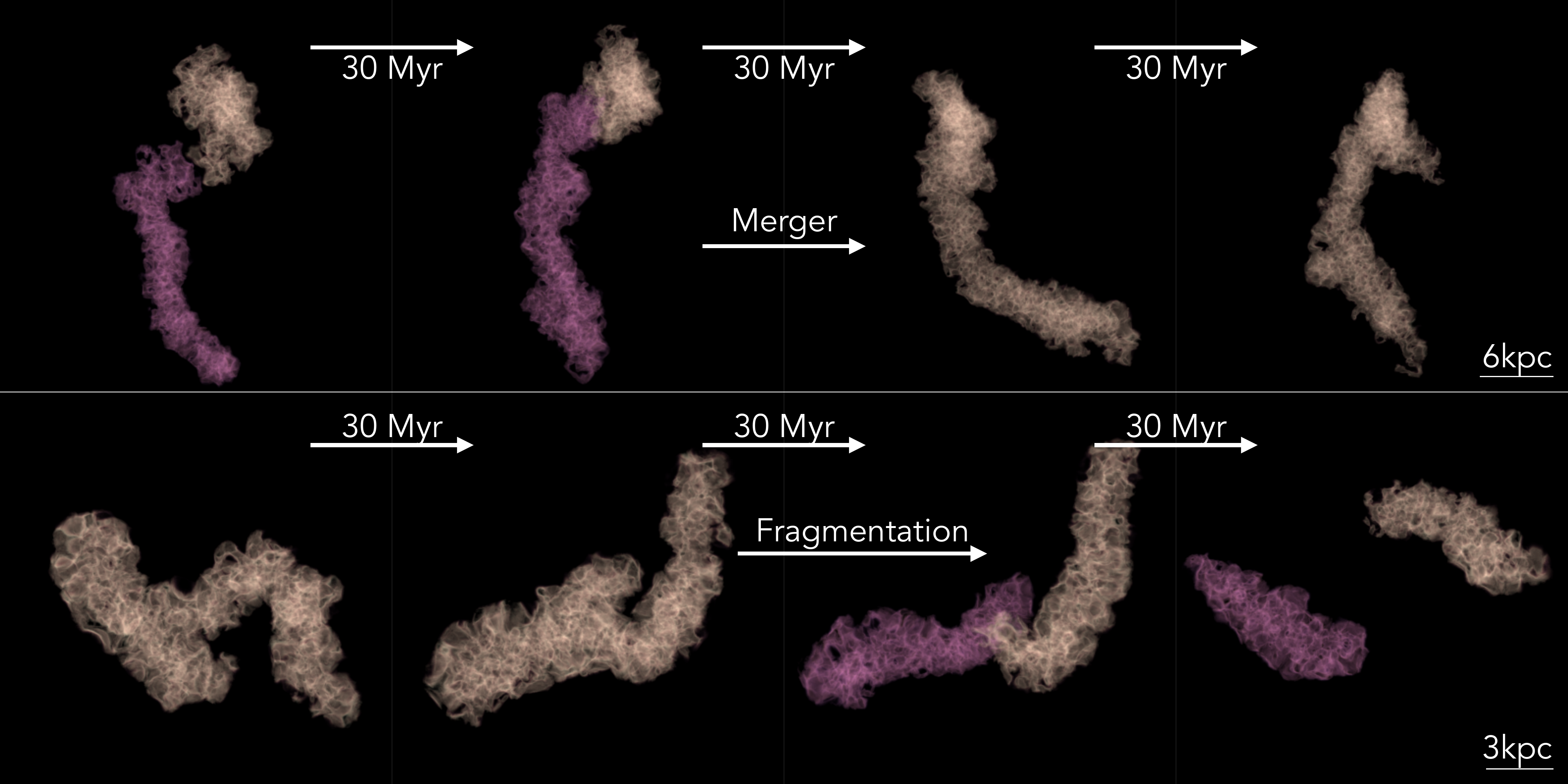}
    \caption{A visualisation of two clouds merging into one (top row) and of a cloud fragmenting into two smaller objects (bottom row). In both rows, neighbouring panels are separated in time by $\sim$\,$30$\,Myr. The scale bars of all panels of a given row are the same, shown in the lower right of each. All panels are oriented such that the mean velocity vector of the cloud (system) is pointed towards the right. Panels with two colors show times when two distinct clouds are present, while panels with only one color contain a single cloud.}
    \label{fig:eventVis}
\end{figure*}

\begin{figure}[ht!]
    \centering
    \includegraphics[width=9cm]{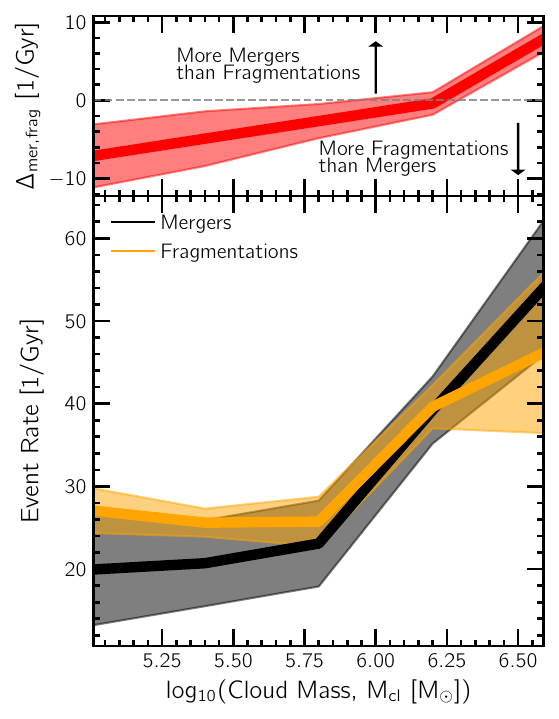}
    \caption{Main panel: Merger and fragmentation rates as a function of ($z$\,$=$\,$0$) cloud mass. The former are shown in black, and the latter in orange. A clear trend of event rates with cloud mass is visible, with lower rates for smaller clouds. In the top panel, we show the difference in rates between merger and fragmentation events: on average, massive clouds merge more often than they fragment, while the opposite is true for their less massive counterparts.}
    \label{fig:cldMergerRate}
\end{figure}

\begin{figure*}[ht!]
    \centering
    \includegraphics[width=18cm]{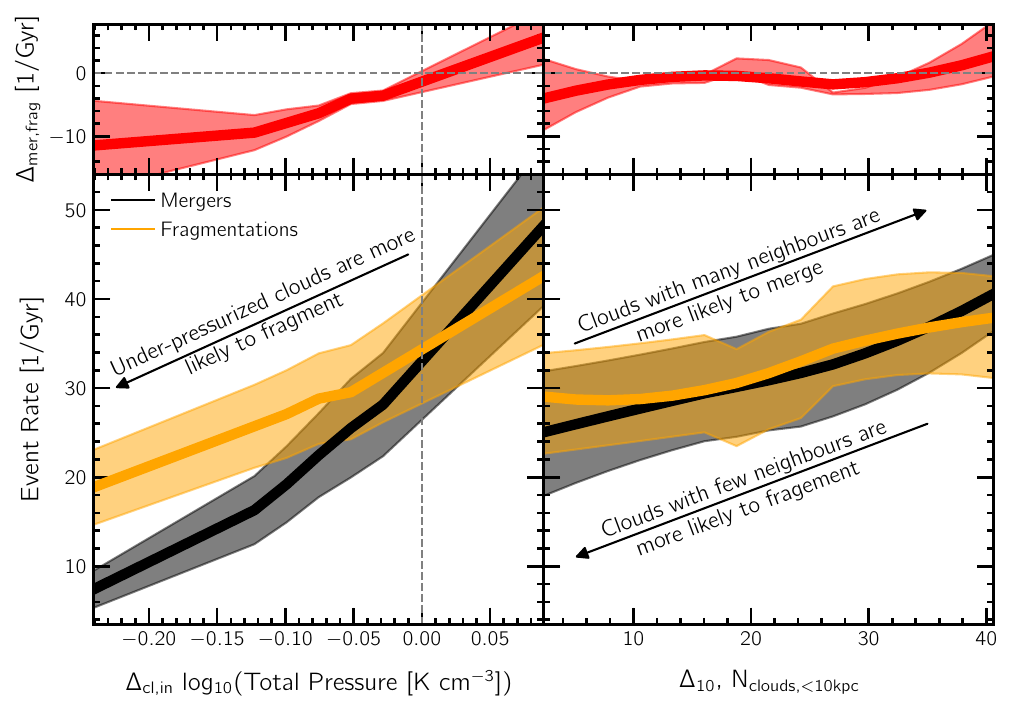}
    \caption{Main panels: merger and fragmentation rates as a function of pressure contrast between the cloud and the interface (left) and $\Delta_{10}$ (right). Top panels: difference between merger and fragmentation rates. A strong trend with pressure contrast is seen, with under-pressurized clouds more likely to undergo a fragmentation event as opposed to a merger. The trend with $\Delta_{10}$ is weaker -- clouds with a very small (large) number of neighbours are likely to fragment (merge), while the net rate is largely flat for intermediate values of $\Delta_{10}$.}
    \label{fig:eventRate_vs_prop}
\end{figure*}

\begin{figure}[ht!]
    \centering
    \includegraphics[width=9cm]{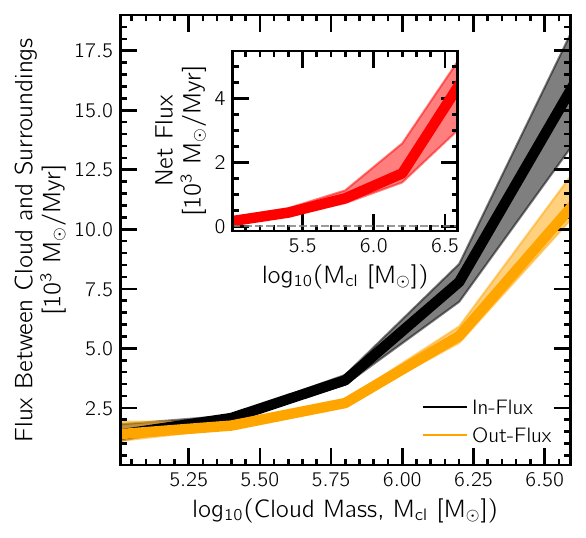}
    \caption{Main panel: Mass in- and out-flux as a function of ($z$\,$=$\,$0$) cloud mass. The former are shown in black, while the latter in orange. Both rates correlate with cloud mass, and are higher for more massive clouds. In the inset, we show the difference between the two rates: the net flux. On average, clouds of all masses experience a net inflow of cold gas into them, with a higher net positive rate for more massive clouds.}
    \label{fig:fluxRate}
\end{figure}

\begin{figure*}[ht!]
    \centering
    \includegraphics[width=18cm]{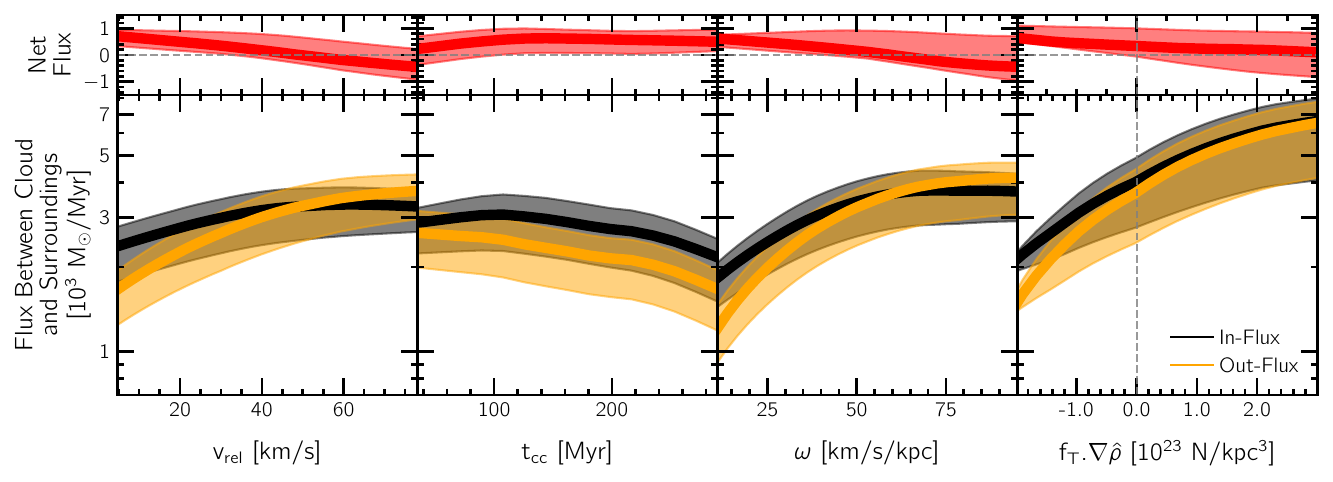}
    \caption{Main panels: in-flux (black) and out-flux (orange) rates as a function of velocity contrast between the cloud and the interface (v$_{\rm{rel}}$; left), cloud crushing time (t$_{\rm{cc}}$; centre-left), vorticity in the interface ($\omega$; centre-right) and the dot product between the magnetic tension vector and density gradient in the interface, i.e. the effective tension (f$_{\rm{T,eff}}$\,$=$\,f$_{\rm{T}}$\,$\cdot$\,$\nabla\hat{\rho}$; right). The top panel shows the difference between the two, i.e. the net mass flow rate. While clouds with relatively low values of v$_{\rm{rel}}$ and $\omega$ experience a net in-flux of material, larger values of these properties trigger a net loss of cold gas from the cloud. The inverse trend holds for t$_{\rm{cc}}$, as higher values correspond to larger shear times, thereby leading to a lower fraction of mass loss from the cloud. The strength and topology of magnetic fields may further be important: a net magnetic tension force acting against the density gradient, i.e. away from the cloud and into the background, results in a net flux of mass into the cloud, likely by inhibiting the mixing of the background into the cloud.}
    \label{fig:fluxRate_vs_prop}
\end{figure*}

\begin{figure}[ht!]
    \centering
    \includegraphics[width=8.8cm]{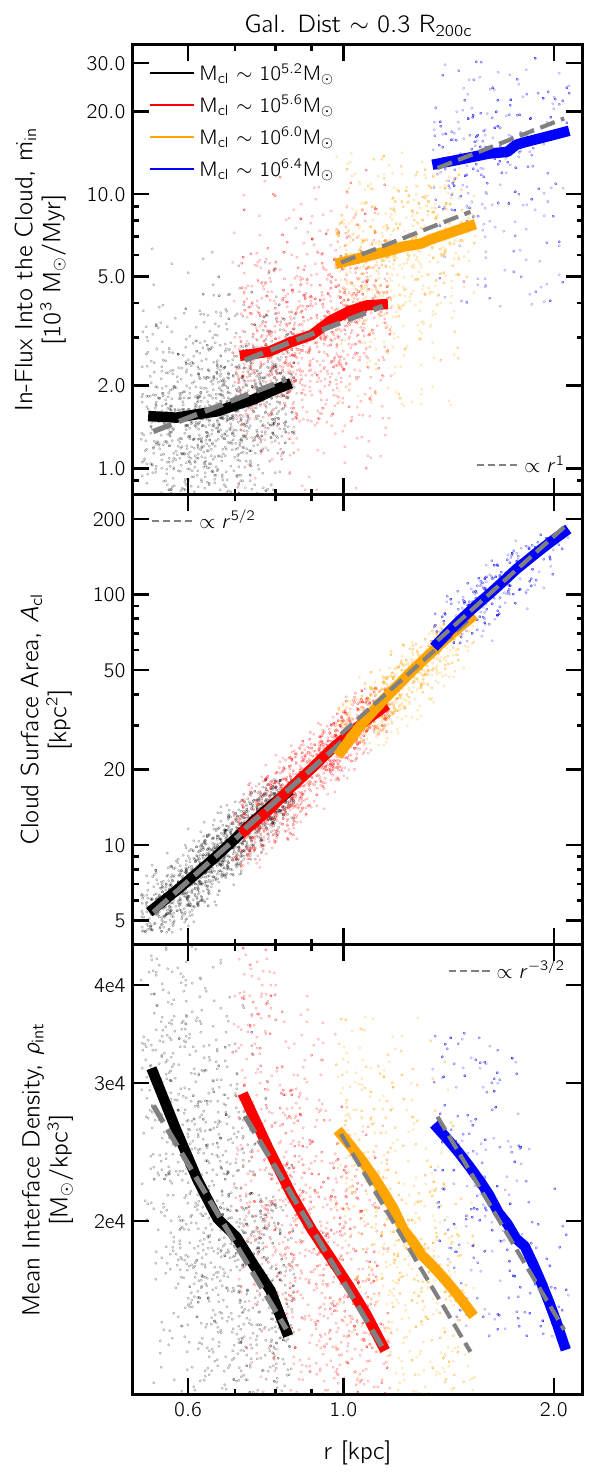}
    \caption{Mass in-flux (top panel), cloud surface area (centre) and mean interface density (lower) as a function of radius, for clouds in the inner halo (galactocentric distance $\sim$ 0.3\,$\rm{R_{200c}}$. The various solid colored curves correspond to median relations for clouds segregated into different mass bins, scatter points to individual clouds, and the dashed gray lines to relevant scaling relations. See main text for more discussion.}
    \label{fig:fluxRate_theory_comp}
\end{figure}

\begin{figure*}[ht!]
    \centering
    \includegraphics[width=18cm]{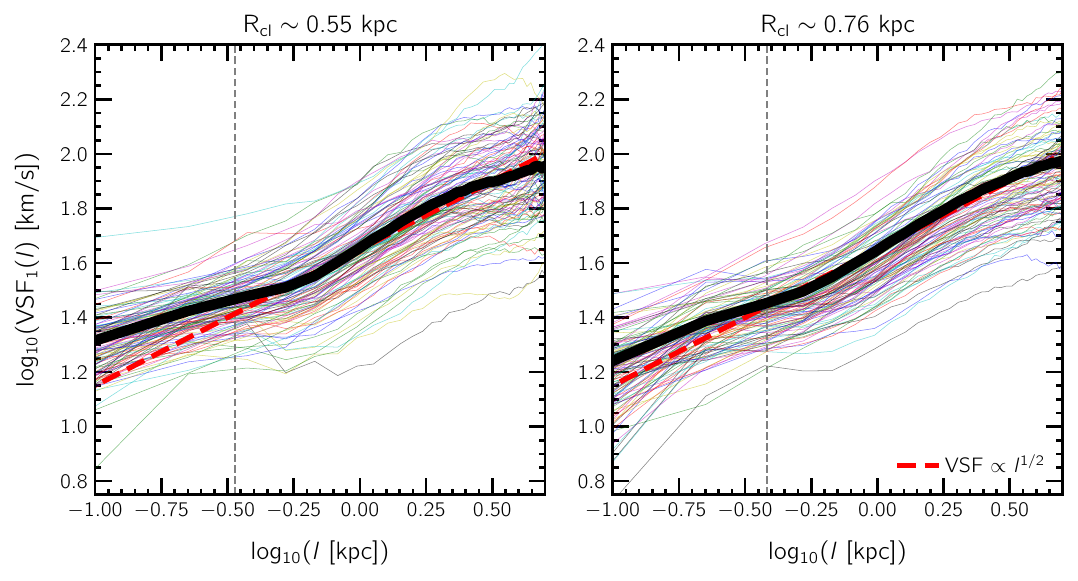}
    \caption{Velocity structure functions (VSFs) for a subset of M$_{\rm cl}$\,$\sim$\,$10^{5.2}$\,M$_\odot$ clouds in the inner halo (galactocentric distance $\sim$ $0.3$\,R$_{\rm 200c}$). The extreme 10$^{\rm{th}}$ percentile of clouds are picked based on cloud radius, with the VSFs of smaller (larger) clouds shown on the left (right). Thin colored curves correspond to individual clouds, thick solid curves to median, and the vertical dashed gray line to the average cell size. Both medians, and a vast majority of the individual curves follow a $\propto$\,$l^{1/2}$ scaling, suggesting that the power law slope of the VSF is $k$\,$\sim$\,$1/2.$}
    \label{fig:vsf_clouds}
\end{figure*}

\begin{figure}[ht!]
    \centering
    \includegraphics[width=9cm]{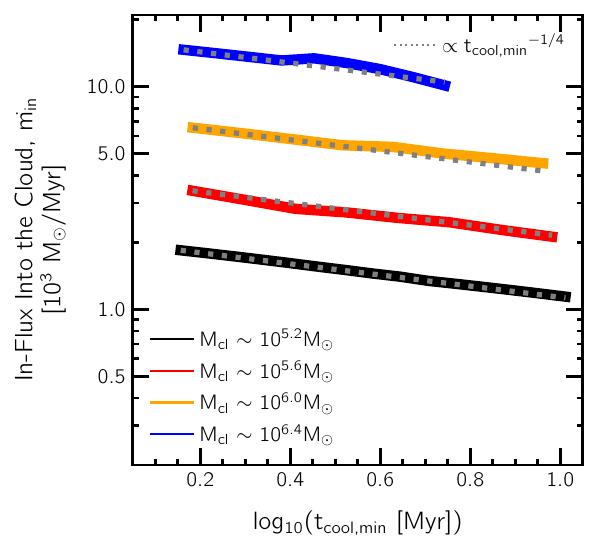}
    \caption{Mass in-flux as a function of $t_{\rm{cool,min}}$ for clouds in the inner halo (galactocentric distance $\sim$\,$0.3$\,$\rm{R_{200c}}$). The colored curves correspond the medians of different mass bins, and the gray curves to relevant scaling relations. The mass in-flux anti-correlates with the cooling time, with a typical power law slope of $\sim$\,$-1/4$, suggesting that these clouds are in the fast cooling regime.}
    \label{fig:fluxRate_t_cool}
\end{figure}

\subsection{Evolution of Long Lived Clouds}\label{ssec:evl_long_lived}

Throughout the rest of this paper, we focus exclusively on the long-lived clouds in our sample. In addition, to limit the stochastic noise associated with Monte Carlo tracers while estimating quantities such as mass flux, we only consider clouds above a mass threshold of $10^{4.8}$\,M$_\odot$. In effect, we restrict the analysis to relatively large clouds, typically resolved by $\gtrsim$\,50 cells and containing $\gtrsim$\,200 tracers. We examine their evolution over the last $\sim$\,$2$\,Gyr, a span of time accessible due to the fully cosmological nature of our simulations.

In Fig.~\ref{fig:evlVis}, we begin with a visualisation of cold clouds in a GIBLE halo at $z$\,$=$\,$0$. The background image shows a ray-traced volume rendering of CGM gas clouds, with the virial radius (R$_{\rm{200c}}$\,$\sim$\,$270$\,kpc) indicated by the white circle \citep[using a visualization method similar to][]{nelson2016}. The various colored surfaces highlight isothermal contours sampled by five narrow Gaussian transfer functions at log$_{10}$(T [K])\,=\,[4.05, 4.15, 4.25, 4.35, 4.45] with a standard deviation of $\sigma$\,$=$\,$0.006$ in log$_{10}$(K), corresponding to different layers of cold gas within clouds. A temperature gradient into the centre of clouds is visible in most cases, with the innermost region typically being the coldest (bluish colors), as opposed to outer layers of clouds that are relatively warmer (reddish colors). Although not seen here, this temperature gradient also extends outwards, i.e. these clouds are surrounded by a warm-interface layer that is sandwiched between the cloud and the hot background, as previously shown by \cite{nelson2020} and \cite{ramesh2023b} for clouds in TNG50. This structure is similar to idealised simulations run at much higher numerical resolution \citep[e.g.][]{gronke2018,fielding2020}, and is believed to play a role in cloud growth and survival by facilitating rapid condensation onto the cloud through radiative cooling \citep{dutta2022}.

In the foreground, we show the (past) trajectory of a $\sim$\,$10^{6.1}$\,M$_\odot$ cloud identified at $z$\,$=$\,$0$, with the various translucent stars highlighting the positions of the main progenitor of this cloud at $25$ distinct snapshots up to $z$\,$\sim$\,$0.15$, using the merger tree constructed as detailed in Section~\ref{ssec:merger_tree}. About $\sim$\,$2$\,Gyr ago (i.e. $z$\,$\sim$\,$0.15$), this cloud is situated close to the virial radius of the halo, and soon begins infalling towards the centre. During this time, the cloud changes significantly in terms of shape, size and mass as it interacts with its surroundings as well as other clouds. This is highlighted in the insets at the corners of the image, which show volume renderings of gas that belongs to the cloud at four distinct snapshots, going back in time (clockwise from the bottom left), with the cloud oriented such that its mean velocity vector points to the right in each case. The non-monotonic evolution of the mass of the cloud is shown by the labels in the corresponding insets, in units of M$_\odot$. Also visible in the top-right inset is an example of an ongoing merger between two clouds, a phenomenon that will be explored and quantified in subsequent parts of this paper. We also highlight the elongated, filamentary, and in general non-spherical geometry of this cloud throughout its evolution, a common feature of cool clouds in GIBLE halos.

Generalizing beyond this single example, Fig.~\ref{fig:cldMass_evl} shows the average evolutionary history of cloud masses over the last $\sim$\,$2$\,Gyr for a population of clouds. As before, we stack clouds across all eight GIBLE halos. In the main panel, we segregate clouds into four bins of width $0.4$\,dex centered on M$_{\rm{cl}}$\,$\sim$\,[$10^{5.0}$, $10^{5.5}$, $10^{6.0}$, $10^{6.5}$]\,M$_\odot$, shown in black, red, orange and blue, respectively. Solid curves show the medians, while the shaded bands correspond to 16$^{\rm{th}}$-84$^{\rm{th}}$ percentile regions. While the four different median curves portray different behaviours up to a lookback time of around a Gyr, they begin to converge at earlier times, i.e. on average, the progenitors of clouds from all four mass bins had roughly the same mass at some point in the past. Specifically, on average, all progenitors were roughly $\sim$\,$10^{6.25}$\,M$_\odot$ in mass about $2$\,Gyr ago. As we elaborate further in subsequent sections, this mass threshold is important with respect to merger and fragmentation events of clouds, and thus to their growth (see also discussion below for resolution dependence). In all cases, the percentile regions are relatively broad: starting from a spread of $\sim$\,$0.3$\,dex at $z$\,$=$\,$0$, the percentile regions are as large as $\gtrsim$\,$2.0$\,dex at earlier times, suggesting that evolutionary tracks of individual clouds are diverse.

In the inset of the main panel, we assess the impact of numerical resolution. Results from the RF512 suite (solid curves) are contrasted against the lower resolution RF64 (dashed) and RF8 (dotted) runs. Curves in the inset are limited to clouds of ($z$\,$=$\,$0$) masses $\sim$\,$10^{6}$ and $10^{6.5}$\,M$_\odot$. Both mass bins converge well at lower resolution. However, the mass at which curves converge is higher for lower resolution runs. Results are thus qualitatively similar across resolution levels, albeit quantitatively different, likely due to the absence of `small' clouds at coarser resolution, thus effectively dragging the median to larger values.

To explore the possible diversity in evolutionary tracks of clouds further, in the top panel, we show the individual histories of a subset of ($z$\,$=$\,$0$) $10^{6.0}$\,M$_\odot$ clouds, split into three categories: those whose masses are overall increasing (left), overall constant (centre) and overall decreasing (right) over the last $2$\,Gyr. We define the first (third) category to be clouds whose progenitors always have a mass lower (greater) than their $z$\,$=$\,$0$ value. For clouds in the second category, we consider those whose progenitors have a mass within $\pm$\,$0.75$\,dex of the $z$\,$=$\,$0$ value. The red dashed line towards the bottom of each of these panels signifies ten times the average baryonic resolution, i.e. the lower limit for the mass of clouds. By definition, individual tracks can thus not extend below this red line.

Evolutionary tracks are clearly unique in each case: clouds evolve through distinct pathways over time, with impulsive responses to merger events similar to galaxies \citep[e.g.][]{rg2016}. While smooth mass changes occur during certain intervals of time, sudden jumps in mass are also visible. To understand this better, for one curve in each panel, we show instances of mergers and fragmentations as identified during the construction of our cloud merger trees. The former are shown in blue dots, the latter in red, and overlapping dots of both colors correspond to instances where both these events take place between neighbouring snapshots. As expected intuitively, sudden jumps in curves are associated with one of these two events, or in some cases, both. The mass growth/loss of clouds thus cannot be studied in isolation, if it is dominated (or significantly affected) by interactions at the cloud population level, as we explore further below.

\vspace{-0.5cm}

\textit{\subsubsection{Merger and Fragmentation Events}}\label{sssec:merg_frag}

Episodes of mergers and fragmentations clearly seem to play an important role in how clouds evolve over time. Idealised cloud-crushing simulations can assess how a cloud fragments into smaller cloudlets over time \citep{klein1994}, and how these fragments may thereafter coagulate to form a larger mass \citep{banda2020,banda2021}. However, mergers between halo-scale distinct clouds are not captured in idealized setups. In this sub-section, we quantify the merger and fragmentation rates expected when a cosmological population of clouds interact with each other, and assess the physical conditions that drive these events.

In Fig.~\ref{fig:eventVis}, we begin with a visualisation of a merger (top row) and fragmentation (bottom row) event. In both cases, neighbouring panels are separated in time by $\sim$\,$30$\,Myr. The scale bars of all panels in a given row are the same, as shown in the bottom-right corner of each row. All panels are oriented with the mean velocity vector of the cloud (system) pointed towards the right. Panels with two colors show times when two distinct clouds exist, while panels with only one color contain a single cloud.

In the top row, we see a case of two clouds moving towards each other, merging into a larger mass, and comoving as a single object thereafter. On the contrary, the bottom panel highlights an example of a single cloud spontaneously breaking into two smaller objects, which drift apart from each other soon after. Note that for simplicity and clarity, we here pick examples where the cloud system is relatively isolated, i.e. these clouds do not interact with any other clouds in this time frame. However, in general, between two snapshots, a cloud may fragment into multiple objects, or multiple clouds may merge together, or both a fragmentation and merger could take place (the latter already seen in Fig.~\ref{fig:cldMass_evl}). Such cases are probably due to limited time resolution between snapshots, and it is likely that cloud interactions take place in a series of binary steps (Fig.~\ref{fig:app_cldNumIntrMerg}), much like the case shown in Fig.~\ref{fig:eventVis}.

In the main panel of Fig.~\ref{fig:cldMergerRate}, we quantify the rate at which merger (black) and fragmentation (orange) events occur. Curves show the mean values, while shaded bands correspond to the $\pm$\,1$\sigma$ variation. As these events are short time-scale phenomena and their rates are sensitive to time cadence between snapshots (Fig.~\ref{fig:app_mergRateTimeResl}), we here only derive rates between $z$\,$=$\,$0$ and $\sim$\,$0.02$, i.e. where we have a time resolution of $\sim$\,$30$\,Myr, and compute mean rates averaged over these snapshots. However, we find no evidence for a trend of events rates with redshift, and thus results from this analysis can likely also be extended out to $z$\,$\sim$\,$0.15$. 

Both merger and fragmentation rates show a clear trend with cloud mass. For instance, a merger rate of $\sim$\,$20$\,Gyr$^{-1}$ for clouds of mass $10^{5.0}$\,M$_\odot$ increases gradually to $\sim$\,$54$\,Gyr$^{-1}$ for their more massive counterparts with $\sim 10^{6.5}$\,M$_\odot$. A similar trend is observed for fragmentation events: a rate of $\sim$\,$26$\,Gyr$^{-1}$ for low mass clouds ($10^{5.0}$\,M$_\odot$) increases to $\sim$\,$46$\,Gyr$^{-1}$ for clouds of mass $10^{6.5}$\,M$_\odot$. However, as mentioned above, event rates are sensitive to time resolution: these numbers are likely an underestimate, and values may be larger by a factor of $\sim$\,4 at an improved snapshot cadence of $\lesssim$\,$6.0$\,Myr (Fig.~\ref{fig:app_mergRateTimeResl}). Regardless, the qualitative mass trends shown here remain unchanged at even higher time resolution.

In the top panel, we show the difference between merger and fragmentation rates ($\Delta_{\rm{mer,frag}}$; y-axis) as a function of $z$\,$=$\,$0$ cloud mass (x-axis). The dashed horizontal line at a value of zero demarcates the regime where mergers are more frequent than fragmentations (above the line) versus the opposite. A net rate of $\sim$\,$-7$\,Gyr$^{-1}$ for clouds of mass $10^{5.0}$\,M$_\odot$ (i.e. 7 fragmentations more than mergers every Gyr) increases to $\sim$\,$8$\,Gyr$^{-1}$ for more massive clouds at $10^{6.5}$\,M$_\odot$ (i.e. 8 mergers more than fragmentations every Gyr), with the transition between negative and positive values occurring roughly at a mass of 
$\sim$\,$10^{6.2}$\,M$_\odot$. We note that this is close to the mass above which clouds in the RF512 run experience a net growth in mass over the last $\sim$\,2\,Gyr (Fig.~\ref{fig:cldMass_evl}), suggesting that merger and fragmentation events indeed play an important role in cloud mass growth over time.\footnote{We note that these results are qualitatively consistent at lower numerical mass resolution. Similar to Fig.~\ref{fig:cldMass_evl}, the red curve in the top panel is shifted further to higher masses with decreasing numerical resolution.} In general, massive clouds (well above the resolution limit) experience a greater number of mergers than fragmentations, and vice versa for their smaller counterparts. The bigger clouds are thus more likely to survive for longer periods of time. We note that similar conclusions have been reached by studies with idealised setups, postulating that clouds above a critical threshold mass have greater chances of increased survivability \citep[e.g.][]{gronke2022}.

In Fig.~\ref{fig:eventRate_vs_prop}, we correlate these event rates with two important physical quantities: local pressure, and a local measure of environmental cloud density i.e. neighbor number. Similar to Fig.~\ref{fig:cldMergerRate}, we show the rates of mergers (black) and fragmentations (orange) in the main panels, while the difference between the two ($\Delta_{\rm{mer,frag}}$) is shown in top panels. Solid curves show the average values, while the shaded regions correspond to the $\pm$\.1$\sigma$ variation at fixed x-axis value. Unlike Fig.~\ref{fig:cldMergerRate} where the event rates were averaged over the last $\sim$\,$300$\,Myr (i.e. the last 10 snapshots), the rates here correspond to those between neighbouring snapshots, and properties are computed at the snapshot preceding the event. This enables us to answer the question: does a given property of the cloud at the present time influences the chances of a merger and/or fragmentation in the near future? While we show results stacked over clouds of all masses, we mention that the following trends are qualitatively similar at roughly fixed cloud mass, although typically steeper for more massive clouds.

We begin by exploring the rate of interactions as a function of the total (i.e. thermal plus magnetic) pressure contrast of the cloud with respect to the interface layer $\Delta_{\rm{cl,in}}$. Positive (negative) values, i.e. those that fall to the right (left) of the vertical line marked at a value of zero, correspond to clouds that are over (under)-pressurised with respect to their local interfaces.  

An average merger rate of $\sim$\,$8$\,Gyr$^{-1}$ for clouds that are heavily under-pressurised increases to $\sim$\,$48$\,Gyr$^{-1}$ for those that are over-pressurised with respect to their interface. While the fragmentation rate also portrays the same qualitative trend, the rise is less steep: a rate of $\sim$\,$19$\,Gyr$^{-1}$ increases to $\sim$\,$42$\,Gyr$^{-1}$ between the two extremities of this plot. The difference between the two, shown in the top panel, shows a clear trend: the net difference between merger and fragmentation events is negative (positive) when clouds are under (over)-pressurised, i.e, these clouds are more (less) likely to fragment that to merge, with the cross-over between the two occurring roughly at a pressure contrast of zero. The pressure contrast is thus an important quantity that has an impact of the occurrence of such interaction events, and thus on the overall growth of clouds \citep{li2020}. Earlier work with TNG50 has shown that clouds are typically severely thermally under-pressurised with respect to their ambient media, and the addition of magnetic pressure brings them closer to pressure balance with their surroundings \citep{ramesh2023b}, highlighting the importance of magnetic fields.

In the right panel, we show trends with respect to $\Delta_{10}$, a proxy for the clustering of clouds around other clouds \citep{ramesh2023b}. It is defined as the number of neighbouring clouds that lie within a sphere of radius 10kpc centred on a given cloud. A value of zero thus implies that there are no neighbouring clouds within this sphere, a value of one corresponds to one neighbour, and so on. Note that, in this approach, we treat clouds as point-like objects located at their centre-of-mass, which is a decent approximation since extended clouds are not as frequent as their smaller counterparts \citep{nelson2020,ramesh2023b,ramesh2024a}. While we only show trends with respect to $\Delta_{10}$ here, we mention that results are qualitatively similar for $\Delta_{20}$ and $\Delta_{30}$.

While both the merger and fragmentation rates increase towards larger values of $\Delta_{10}$, the trend is weak. An average merger rate of $\sim$\,$26$\,Gyr$^{-1}$ at $\Delta_{10}$\,$\sim$\,$3$ increases gradually to $\sim$\,$40$\,Gyr$^{-1}$ at $\Delta_{10}$\,$\sim$\,$40$. The rate of fragmentation events is rather similar, with slightly elevated (decreased) rates at low (high) $\Delta_{10}$, as seen in the top panel: the difference between the two rates is negligible for most values of $\Delta_{10}$ (i.e. $10$\,$\lesssim$\,$\Delta_{10}$\,$\lesssim$\,$35$). However, at lower values, the net rate is negative, signifying that fragmentation events are more common than mergers when clouds are surrounded by very few counterparts, as intuitively expected. On the other extreme, when clouds are strongly clustered and have a larger number of neighbours, the merger rate overtakes that of fragmentations. In this case clouds have a net positive mass growth. The spatial clustering of clouds may thus play more than one important role in cloud evolution: regulating the net merger rate and thus mass growth, while also increasing cloud lifetimes through the process of drafting, as seen in wind tunnel setups \citep{williams2022}. 

\textit{\subsubsection{Mass Flux Between Clouds and their Surroundings}}\label{sssec:flux}

Results from various idealised simulations suggest that the mass flux into and out of clouds, and their net balance, is an important factor that governs cloud growth and survivability. Fluid instabilities such as the Kelvin-Helmholtz instability trigger mixing of the cloud with the ambient gas, thereby driving an outflux from the cloud \citep{agertz2007}. This mixing gives rise to an intermediate layer of warm gas which, if radiative cooling is substantial, may condense down onto the cloud resulting in an influx of cold gas \citep[e.g.][]{melioli2005,marinacci2010,armillotta2016,gronke2018}. In this sub-section, we quantify the in- and out-flux rates of CGM clouds.

In Fig.~\ref{fig:fluxRate}, we show the flux rates as a function of cloud mass at $z$\,$=$\,$0$ for the same sample of long-lived clouds as above. Similar to merger rates, we find that flux rates are sensitive to the time cadence between snapshots, and hence restrict the computation of fluxes to the final 10 snapshots ($0.0$\,$\lesssim$\,$z$\,$\lesssim$\,$0.02$) which have a time resolution of $\sim$\,$30$\,Myr. As before, fluxes are averaged over these 10 snapshots ($\sim$\,$300$\,Myr). However, we note that we find no strong trend of flux rates with redshift, and the results of this analysis can likely be extended out to $z$\,$\sim$\,$0.15$. Solid curves show the median, while the shaded bands correspond to the 16$^{\rm{th}}$-84$^{\rm{th}}$ percentile regions.

The main panel shows the mass flux rates into (black) and out of (orange) clouds. While both are roughly $\sim$\,$1.5 \times 10^3$\,M$_\odot$\,Myr$^{-1}$ for clouds of mass $\sim$\,$10^5$\,M$_\odot$, the in-flux increasing sharply by a factor of around ten to $\sim$\,$15 \times 10^3$\,M$_\odot$\,Myr$^{-1}$ for clouds of mass $\sim$\,$10^{6.5}$\,M$_\odot$. While the out-flux also scales with mass, the rise is less steep, with an average out-flux of $\sim$\,$11 \times 10^3$\,M$_\odot$\,Myr$^{-1}$ for clouds of mass $\sim$\,$10^{6.5}$\,M$_\odot$. 

The difference between the two, i.e. the net flux, is shown in the inset. Interestingly, on average, the net rate is positive for all mass clouds we study: a flux of $\sim$\,$0.15 \times 10^3$\,M$_\odot$\,Myr$^{-1}$ for low mass clouds ($\sim$\,$10^5$\,M$_\odot$) increases to $\sim$\,$4.5 \times 10^3$\,M$_\odot$\,Myr$^{-1}$ for their more massive counterparts ($\sim$\,$10^{6.5}$\,M$_\odot$). While the net flux averaged over $\sim$\,$300$\,Myr$^{-1}$ is positive, note that there may be clouds that experience a net out-flow of mass between snapshots, as we will shortly explore. In conjunction with Fig.~\ref{fig:cldMass_evl}, this result suggests that, for small clouds, the in-flow of cold material has a sub-dominant impact on mass growth in comparison to cloud-cloud interactions: despite the (small) net in-flow of gas, their overall mass decreases as a result of repeated fragmentations.

In Fig.~\ref{fig:fluxRate_vs_prop}, we explore four important physical properties that impact the mass flux into and out of clouds. The main panels shows the in-flux (black) and out-flux (orange) rates. Solid curves correspond to the median, and the top panels show the difference between the two, i.e. the net mass flux. The horizontal dashed gray line demarcates the regime of net in-flow into the cloud (above the line, positive values) from that of a net out-flow. Similar to Fig.~\ref{fig:eventRate_vs_prop}, the rates here correspond to those between neighbouring snapshots, and properties are computed at the snapshot preceding the mass exchange. As before, while we show results stacked over clouds of all masses, we mention that the following trends are qualitatively similar at roughly fixed cloud mass, although typically steeper for more massive clouds.

The left panel shows flux rates as a function of the velocity contrast between the cloud and the interface (v$_{\rm{rel}}$). An average in-flux of $\sim$\,$2.5 \times 10^3$\,M$_\odot$\,Myr$^{-1}$ for clouds at similar velocities as their interface (v$_{\rm{rel}}$\,$\lesssim$\,$5$\,km\,s$^{-1}$) increases to $\sim$\,$3.25 \times 10^3$\,M$_\odot$\,Myr$^{-1}$ for clouds moving at larger relative velocities (v$_{\rm{rel}}$\,$\sim$\,$80$\,km\,s$^{-1}$). The growth of the out-flux rate is more steep, increasing from $\sim$\,$1.8 \times 10^3$ to $\sim$\,$4.0 \times 10^3$\,M$_\odot$\,Myr$^{-1}$ between the two extremities of this plot. 

On average, the net flux is positive for relatively small velocity contrasts (v$_{\rm{rel}}$\,$\lesssim$\,$40$\,km/s), signifying a net inflow of cold gas into the cloud. For larger contrasts, on average, clouds experience a net loss of mass into the ambient gas. These results are qualitatively consistent with those from higher resolution idealised studies which have argued that a large velocity shear may lead to an overall loss of cloud mass as a result of fluid instabilities such as the KH-instability \citep[e.g.][]{scannapieco2015}, or by resulting in an increase of the turbulent velocity that may further have an impact on cloud growth over time \citep[e.g.][]{abruzzo2024}.

In addition to v$_{\rm{rel}}$, the size ($r$) and the overdensity of the cloud ($\chi$) influence the time-scale over which the cloud is impacted by these fluid instabilities. Commonly referred to as the cloud crushing time-scale, this is defined as t$_{\rm cc}$\,$=$\,$\chi^{1/2}\,r\,\rm{v_{rel}^{-1}}$ \citep{klein1994}. In the centre-left panel, we explore the dependence of the flux rates on t$_{\rm cc}$. While both the in- and out-fluxes anti-correlate with t$_{\rm cc}$, the net-flux increases towards larger t$_{\rm cc}$ and eventually flattens out. We construe such a trend to arise since clouds with larger t$_{\rm cc}$ have larger shearing time scales, thereby leading to a lower fraction of
mass loss from the cloud due to mixing with the background \citep{scannapieco2015}. It is however important to note that the cooling time-scales of the interface may vary across the sample of clouds, which we explore in Section~\ref{ssec:comp_theory}. 

Fluid instabilities drive the mixing of material across the interface, effectively cycling matter between the cloud and the background \citep[e.g.][]{fielding2020}. In the centre-right panel, we aim to quantify the impact of this on flux rates, through measurements of vorticity ($\omega$\,$=$\,$|\Vec{\nabla} \times \Vec{v}|$) in the interface. Trends are similar to the left panel, as expected since v$_{\rm{rel}}$ and $\omega$ correlate strongly with each other \citep{ramesh2024b}: the in-flux rate increases from $\sim$\,$1.9 \times 10^3$ to $\sim$\,$3.9 \times 10^3$\,M$_\odot$\,Myr$^{-1}$ from low ($\sim$\,$10$\,km\,s$^{-1}$\,kpc$^{-1}$) to high values of vorticity ($\sim$\,$95$\,km\,s$^{-1}$\,kpc$^{-1}$). The out-flux rate increases more steeply, from $\sim$\,$1.2 \times 10^3$ to $\sim$\,$4.2 \times 10^3$\,M$_\odot$\,Myr$^{-1}$ over the same range of $\omega$. 

Overall, the net flow rate is positive for small values of $\omega$, corresponding to a net flow of cold gas into the cloud. For large values of $\omega$ the net flow becomes negative, signifying a transfer of mass from the cloud into the ambient gas. We interpret these two limiting cases as arising from a tug-of-war between two key physical processes. As long as the rate of mass loss from the cloud due to mixing is smaller than the rate at which gas from the ambient gas cools and condenses onto the cloud, the cloud mass increases over time \citep[e.g.][]{fielding2022}. 

In addition to radiative cooling, it is believed that non-thermal processes, such as magnetic fields, may be important. For instance, as clouds pass through the halo, they may sweep up ambient field lines to create a draped layer around them \citep[e.g.][]{dursi2008,pfrommer2010}, which may thereafter suppress fluid instabilities along their interfaces \citep[e.g.][]{sparre2020,cottle2020}. We next assess the possible impact that such a phenomenon may have on mass flux rates. We do so by first computing the magnetic tension, $\Vec{\rm{f}_{\rm{T}}}$\,$=$\,$|\Vec{B}|^2$\,$\Vec{\kappa}$, where $\Vec{\kappa} = (\Vec{b} \dcdot \nabla) \Vec{b}$ is the magnetic curvature vector \citep{shen2003,boozer2005}. Here, $\Vec{b}$ = $\Vec{B} / B$ is the unit vector in the direction of the magnetic field $\Vec{B}$. We have previously explored the amount of magnetic draping and its impact on CGM gas clouds  \citep{ramesh2024b}.

Here, we take the dot product of the tension vector with the unit vector of the density gradient $\Vec{\nabla\hat{\rho}}$ to yield an effective tension force f$_{\rm{T,eff}}$\,$=$\,$\Vec{\rm{f}_{\rm{T}}}$\,$\cdot$\,$\Vec{\nabla\hat{\rho}}$. Since the gas density decreases from the cloud center towards its outskirts, and thereafter into the background \citep{nelson2020,ramesh2023b}, $\Vec{\nabla\hat{\rho}}$ points into the cloud from the background. Positive values of f$_{\rm{T,eff}}$ thus correspond to cases where the tension is directed into the cloud, while the opposite holds for negative values.

The right panel of Fig.~\ref{fig:fluxRate_vs_prop} shows the different flux rates as a function of the effective tension, f$_{\rm{T,eff}}$. On average, both the in- and out-flux rates are largely similar for positive values of f$_{\rm{T,eff}}$, i.e. the net mass flux rate is $\sim$\,$0$ in this regime. However, for negative values of f$_{\rm{T,eff}}$, we note a net in-flow of mass into clouds. We interpret this as the result of the tension force inhibiting a flow of gas from the background into the cloud (i.e. along the density gradient), thereby reducing mixing and vorticity, and eventually leading to a net in-flux of material. While draped configurations of magnetic fields are believed to possibly also suppress the impact of thermal conduction \citep{ettori2000,bruggen2023}, which might otherwise contribute to cloud evaporation and destruction \citep{marcolini2005,vieser2007}, the current GIBLE simulations do not include this physical mechanism. Future extensions that include conduction can explore its impact on cloud evolution in a cosmological context.  

\vspace{0.5cm}

\subsection{Comparison with Theoretical Expectations}\label{ssec:comp_theory}

So far we have qualitatively discussed our results in the context of previous work from the literature. We now provide a more quantitative comparison with analytic models and results from idealised simulation studies. In particular, we focus on the theoretical expectations for mass flux rates in to and out of clouds.

In Fig.~\ref{fig:fluxRate_theory_comp}, we show a number of quantities of interest as a function of cloud radius $r$. The various solid colored curves correspond to medians for clouds segregated into different mass bins, scatter points to individual clouds, and the dashed gray
lines to relevant scaling relations. Since the measurement of $r$ is sensitive to the galactocentric distance of the cloud, as this impacts the mean density of gas, we focus on clouds in the inner halo ($\sim$\,$0.3$\,$\rm{R_{200c}}$) where statistics are the best \citep{ramesh2023b}, and mention that these results and discussion hold true for clouds in other regions of the halo as well. As before, we restrict the selection of clouds to the final ten snapshots. Flux rates correspond to those between neighbouring snapshots, and all properties of clouds are computed at the snapshot preceding the mass exchange.

In the top panel, we begin with the mass in-flux into the cloud, $\dot{m_{\rm in}}$. In all four mass bins, $\dot{m_{\rm in}}$ increases gradually towards larger $r$, and is well represented by a $\propto$\,$r^1$ scaling in all cases. In a recent study by \citealt{tan2024}, it was suggested that $\dot{m_{\rm in}}$\,$\propto$\,$r^{11/4}$ for clouds in their tall box setup. Note, however, that this was the cumulative scaling for all clouds, i.e. not split by mass. In our simulations, we find that M$_{\rm{cl}}$\,$\propto$\,$r^{3/2}$ (not shown), implying a $\dot{m_{\rm in}}$\,$\propto$\,$r^{5/2}$ scaling if clouds of all masses are stacked together, yielding a largely similar power law slope as reported by \cite{tan2024}.

To understand the $\dot{m_{\rm in}}$\,$\propto$\,$r^1$ scaling further, we consider a model for cloud growth where the in-flux into the cloud ($\dot{m_{\rm in}}$) is a function of the surface area $A_{\rm{cl}}$, density of the interface $\rho_{\rm{int}}$, and the inflow velocity of gas into the cloud $v_{\rm{in}}$ (see \citealt{gronke2020, tan2024}):

\begin{equation}
    \dot{m_{\rm in}} \sim A_{\rm{cl}} \cdot  \rho_{\rm{int}} \cdot v_{\rm{in}} 
\label{eq:in_flux}    
\end{equation}

The centre panel of Fig.~\ref{fig:fluxRate_theory_comp} shows $A_{\rm{cl}}$ as a function of $r$. All curves follow a $\propto$\,$r^{5/2}$ scaling, with a tight correlation of the scatter points around their respective medians. This is in remarkable agreement with results from idealised setups run at much higher resolution that have shown how the surface areas of clouds that continuously experience a shear force scale as $A_{\rm{cl}}$\,$\propto$\,$r^{5/2}$ \citep{gronke2023,tan2023}, corresponding to a fractal dimension $D$\,$=$\,$2.5$ in the mixing layer \citep{fielding2020}.

The variation of $\rho_{\rm{int}}$ with $r$, shown in the bottom panel, is non-trivial. While it is typically assumed that $\rho_{\rm{int}}$ is independent of $r$, we find a reasonably strong $\rho_{\rm{int}}$\,$\propto$\,$r^{-3/2}$ scaling in all four mass bins. While not shown here, we mention that $\rho_{\rm{cloud}}$, the mean density of the cloud, also is $\propto$\,$r^{-3/2}$, and $\rho_{\rm{int}}$ inherits this scaling as clouds aim to maintain pressure balance with their surroundings. We find that these trends are driven by the inhomogeneities of the CGM at fixed galactocentric distance. Such large pressure fluctuations imply that clouds embedded in low pressure environments have lower densities, and thus larger radii at fixed mass \citep[see also][]{stern2021}.

Radiative turbulent mixing layer simulations suggest that the final ingredient of the model, $v_{\rm{in}}$, can be written as \citep{fielding2020}:

\begin{equation}
    v_{\rm{in}} \sim  t_{\rm{cool, min}}^{-(1-\frac{1}{2(1-k)}))} \cdot L_{\rm{turb}}^{1-\frac{1}{2(1-k)}} \cdot v_{\rm{turb, L}}^{\frac{1}{2(1-k)}}
\label{eq:v_in}    
\end{equation}

where $t_{\rm{cool, min}}$ is the minimum cooling time in the interface layer, $L_{\rm{turb}}$ is the outer scale of turbulence, and $v_{\rm{turb, L}}$ is the turbulent velocity at large scales, i.e. the typical length scale along of turbulent driving. Lastly, $k$ is the power-law slope of the velocity structure function (VSF), i.e. VSF\,$\propto$\,$l^k$, where $l$ is the separation scale. To compute $v_{\rm{in}}$, one thus first needs to estimate $k$.

In Fig.~\ref{fig:vsf_clouds}, we show VSFs for a subset of clouds from the M$_{\rm cl}$\,$\sim$\,$10^{5.2}$\,M$_\odot$ bin. In particular, to probe possible trends as a function of $r$, we pick the extreme 10$^{\rm{th}}$ percentiles of clouds based on size. VSFs of the smaller (larger) set of clouds are shown on the left (right). Thin colored curves correspond to VSFs of individual clouds, while the thick black curve to the median. We compute the VSF with the first three layers of Voronoi cells around clouds, i.e. the interface plus two further background layers, although we have verified that the following results are not sensitive to this exact choice. The vertical dashed gray curves corresponds to the average size of the cells selected for the analysis.

To guide the eye, the red dashed curves show a $\propto$\,$l^{1/2}$ scaling. Medians in both panels, as well as most individual curves, seem to closely trace this scaling, suggesting that the power-law index $k$ of the VSFs in the regions around these clouds is $\sim$\,$1/2$. Although not shown, VSFs of other clouds, i.e. other masses and galactocentric radii, show similar behaviour. The VSFs in both panels also appear to flatten out at roughly the same separation scales, implying that $L_{\rm{turb}}$ is largely independent of $r$. This is in agreement with \cite{tan2024}, where the same scaling was suggested for their sample of clouds. 

Note, however, that resolution is limited on these scales, and it is possible that the inferred VSFs are influenced by numerical dissipation \citep[e.g.][]{bauer2012,zier2022}. Comparison with higher resolution cosmological simulations run in the future will help assess and verify the accuracy of these inferences. For the present work, adopting that $k$\,$=$\,$1/2$, Equation~(\ref{eq:v_in}) reduces to $v_{\rm{in}}\,\sim\,v_{\rm{turb, L}}^{1}$, i.e. $v_{\rm{in}}$ in this case depends only on $v_{\rm{turb, L}}$, and is independent of $t_{\rm{cool, min}}$ and $L_{\rm{turb}}$. To estimate the dependence of $v_{\rm{in}}$ on $r$, we need to assess the analogous scaling of $v_{\rm{turb, L}}$.

Measuring $v_{\rm{turb, L}}$ directly from the simulation output requires knowledge of the associated turbulence mixing length scale, $L_{\rm{turb}}$. While $L_{\rm{turb}}$ can be derived from the VSF \citep[see e.g.][]{abruzzo2024}, we instead adopt the simpler approach of comparing the normalisations of the VSFs as a function of $r$. As seen in the two panels of Fig.~\ref{fig:vsf_clouds}, and many other cases that we have tested, this is largely independent of $r$, implying that $v_{\rm{turb, L}}$\,$\propto$\,$r^{0}$.

Plugging in these different scaling relations ($A_{\rm{cl}}$\,$\propto$\,$r^{5/2}$, $\rho_{\rm{int}}$\,$\propto$\,$r^{-3/2}$ and $v_{\rm{in}}$\,$\sim$\,$v_{\rm{turb, L}}$\,$\propto$\,$r^{0}$) into Equation~(\ref{eq:in_flux}) yields $\dot{m_{\rm in}} \sim r^{5/2} \cdot r^{-3/2} \cdot r^{0}$\,$\implies$\,$\dot{m_{\rm in}} \sim r^{1}$, in excellent agreement with the top panel of Fig.~\ref{fig:fluxRate_theory_comp}. These results are consistent with $k$\,$=$\,${1/2}$ and the corresponding limit of the cloud growth model. We note once again, however, that resolution on these scales is limited in our simulations, and a convergence study towards even better numerical resolution will clarify this picture.

As a final point of comparison, we assess the relation between the mass in-flux into the cloud and the minimum cooling time in the interface. Turbulent mixing layer simulations predict the flux to scale as $\propto$\,$t_{\rm{cool,min}}^{-1/4}$ in the fast cooling regime, with a larger negative power law slope ($\propto$\,$t_{\rm{cool,min}}^{-1/2}$) in the regime of slow cooling \citep{ji2018,fielding2020,tan2021}. 

In Fig.~\ref{fig:fluxRate_t_cool}, we examine these trends for clouds in our simulation. As before, we select clouds in the inner halo (galactocentric distance $\sim$\,$0.3$\,$\rm{R_{200c}}$) to excise the radial trend of $t_{\rm{cool,min}}$, and segregate clouds into bins based on their mass. The corresponding medians are shown through colored solid curves, while the dotted gray lines show scalings of $\propto$\,$t_{\rm{cool,min}}^{-1/4}$.

The mass in-flux in all cases anti-correlates with the minimum cooling time, i.e. clouds with faster cooling interfaces experience a larger in-flux of material. The power law-slope in all cases is $\sim$\,$-1/4$, suggesting that these clouds are in the regime of fast cooling. Although not shown explicitly, we find that the net-flux also scales as $t_{\rm{cool,min}}^{-1/4}$ at low-$t_{\rm{cool,min}}$, while the power law slope tends to more negative values at larger $t_{\rm{cool,min}}$. A more quantitative comparison with the mass flux rates expected from theoretical models will be enabled by future simulations.

% ------------------------------------------------------------------------------

\section{Summary}\label{sec:summary}

In this paper we quantify the origin and evolution of cold gas clouds in Project GIBLE, a suite of cosmological magneto-hydrodynamical zoom-in simulations with super-Lagrangian refinement in the CGM down to a mass resolution of $\sim$\,$10^3$\,M$_\odot$. Focusing on halos of Milky Way-like galaxies, we use Monte-Carlo tracers to identify progenitors and descendants of clouds across time. We construct cloud merger (interaction) trees and use these to identify cloud origins and explore the mass growth histories of clouds. Our main findings are as follows:

\begin{enumerate}

    \item The population of $z$\,$=$\,$0$ clouds primarily arises from two roughly balanced origins: recent ($\lesssim$\,$2$\,Gyr) outflows of cold gas driven from the galaxy into the CGM ($\sim$\,40-60\,$\%$) and precipitation of hot halo gas ($\sim$\,25-50\,$\%$). More massive clouds are more likely to arise from outflows, while low-mass clouds are more likely to originate from hot halo cooling. Only $\sim$\,5\,$\%$ of clouds are direct descendants of cold gas stripped or otherwise removed from satellites (Figs.~\ref{fig:originVis} and \ref{fig:originFraction}).

    \item Properties of $z$\,$=$\,$0$ clouds are diverse and depend of their origin. Clouds stripped from satellites are typically smaller in size and situated at larger galactocentric distances. The distribution of cloud metallicity is bimodal, with a primary peak at $\sim$\,$0.45$\,Z$_\odot$, and a smaller secondary peak at $\sim$\,$0.15$\,Z$_\odot$. These features reflect specific accretion histories of massive satellites. In contrast, clouds originating from the central galaxy or hot halo condensation have intermediate metallicities of $\sim$\,$0.25$\,Z$_\odot$ (Fig.~\ref{fig:origin_vs_prop}). 

    \item Clouds that precipitate out of the hot halo begin from a cold overdense seed. Rapid cooling and accretion leads to its growth over time, and such cold seeds eventually reach a mass of $\sim$\,$10^{5}$\,M$_\odot$ by $z$\,$=$\,$0$. This exact value decreases as numerical resolution improves. Cold seeds are primarily created as a result of short time-scale ($\lesssim$\,$500$\,Myr) dissolution of previous cold gas into the hot phase, which produces density fluctuations sufficient for the next generation of clouds to form (Fig.~\ref{fig:cldEvlCon}).
    
    \item The baryonic material that builds cool CGM clouds, as with all baryonic matter in halos, is ultimately of cosmological origin. Roughly $\sim$\,$85$\,$\%$ of their mass comes from smooth accretion from the IGM at earlier cosmic epochs (Fig.~\ref{fig:seedOrigin}).

    \item We make predictions for the origin of cold clouds in the Milky Way halo: while LVCs and IVCs are roughly equally likely to arise as a result of cold galactic outflows and hot-phase precipitation, outflowing (inflowing) HVCs are predominantly signatures of the former (latter). Clouds arising from the cold gas of satellites are rare, but equally likely at all line-of-sight velocities (Fig.~\ref{fig:originFraction_vs_vlos}).

    \item Roughly $\sim 10\%$ of the $z$\,$=$\,$0$ cloud population is long-lived, with their progenitors having already assembled $\sim$\,$2$\,Gyr ago. These clouds have unique and diverse evolutionary tracks as they pass through the halo, interact with other clouds, and exchange mass with their surroundings. Clouds portray a temperature gradient into their surroundings with a warm interface layer separating them from the hot background, which is believed to improve chances of their survivability (Figs.~\ref{fig:evlVis} and \ref{fig:cldMass_evl}).

    \item Long-lived clouds merge and fragment at an average rate of $\gtrsim$\,$20$\,Gyr$^{-1}$. Above some critical mass threshold, more massive clouds undergo more mergers than fragmentations (Fig.~\ref{fig:cldMergerRate}). This results in net mass growth over time. The frequency of these interactions is modulated by the pressure imbalance between clouds and their interfaces, with under-pressurised clouds more likely to fragment. The clustering of clouds around other clouds plays a secondary role: those with many neighbours are more likely to merge (Fig.~\ref{fig:eventRate_vs_prop}).

    \item Clouds also exchange matter with their surrounding gas, at a rate of $\gtrsim$\,$10^{3}$\,M$_\odot$\,Myr$^{-1}$, and massive clouds experience a larger net-flux (Fig.~\ref{fig:fluxRate}). Small-scale physical properties have an impact on mass exchange: the net-flux into the cloud anti-correlates (correlates) with the velocity contrast, as well as interface-layer vorticity (cloud crushing time scale). The topology and magnitude of magnetic fields may further play a role: we find that a net magnetic tension force acting against the density gradient can help suppress mixing of the background into the cloud (Fig.~\ref{fig:fluxRate_vs_prop}). In addition, the flux rates in and out of clouds may also depend on the surface area, as well as the cooling time of gas in the interface (Figs.~\ref{fig:fluxRate_theory_comp} and \ref{fig:fluxRate_t_cool}).
    
\end{enumerate}

This work provides an initial set of cosmological inputs to questions related to the origin and evolution of cool clouds in the circumgalactic medium of galaxies. Clear avenues for future extensions exist. What is the fate of these clouds? Do they fuel the cold gas supply of the central galaxy to sustain star formation, or perish as they accrete towards the centre? If they do increase the galactic cold mass budget, does this impact future outflows, and the production of new clouds? We reserve such explorations for the future.

\begin{acknowledgements}
RR and DN acknowledge funding from the Deutsche Forschungsgemeinschaft (DFG) through an Emmy Noether Research Group (grant number NE 2441/1-1). RR is a Fellow of the International Max Planck Research School for Astronomy and Cosmic Physics at the University of Heidelberg (IMPRS-HD). RR and DN thank Ralf Klessen and Annalisa Pillepich for valuable inputs. MB acknowledges funding by the Deutsche Forschungsgemeinschaft (DFG, German Research Foundation) under Germany's Excellence Strategy -- EXC 2121 ``Quantum Universe'' --  390833306. This work has made use of the VERA supercomputer of the Max Planck Institute for Astronomy (MPIA) operated by the Max Planck Computational Data Facility (MPCDF), and of NASA's Astrophysics Data System Bibliographic Services. 
\end{acknowledgements}

\bibliographystyle{aa}
\bibliography{references}

\appendix
\section{Tracing Cosmic Gas Accretion}\label{app:accrn}

\begin{figure*}
    \centering
    \includegraphics[width=18cm]{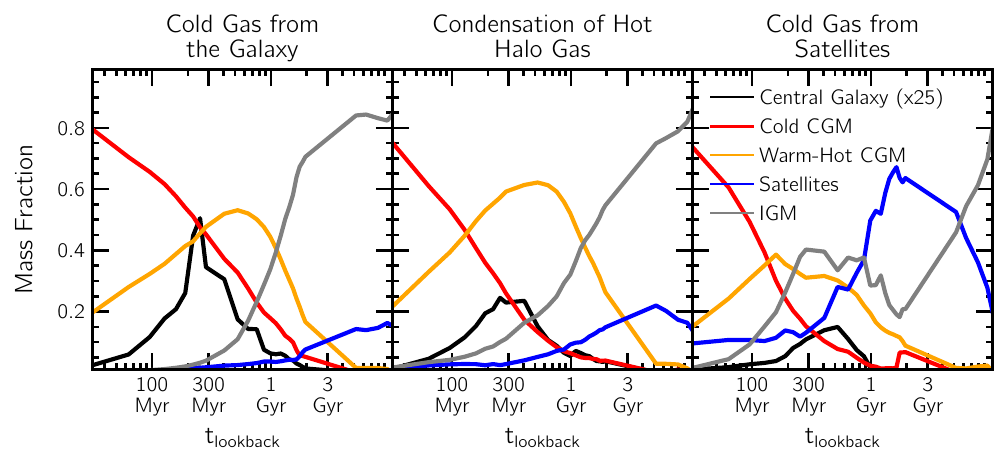}
    \caption{Similar to Fig.~\ref{fig:seedOrigin}, we dissect the various sources of gas that populate $z$\,$=$\,$0$ clouds, split based on their origin source: those arising from the central galaxy in the left panel, through condensation of hot halo gas in the centre, and from satellites in the right. The five different colors correspond to the distinct pathways that we consider. Similar to cold seeds, $\gtrsim$\,$80$\,$\%$ of the $z$\,$=$\,$0$ mass was in the IGM at $z$\,$\sim$\,$2$.}
    \label{fig:app_cloudOriginHighZ}
\end{figure*}

In this section, we extend the analysis of Fig.~\ref{fig:seedOrigin}, dissecting the various sources of gas that populate all clouds at $z$\,$=$\,$0$, not just seeds as previously explored. We consider five distinct sources: (a) gas from the central galaxy (all phases of gas included) shown in black, (b) the cold CGM, i.e. clouds that existed in the past but dissolved into the hot phase, in red, (c) the warm-hot CGM in orange, (d) gas from satellite galaxies (all phases included) in blue, and (e) the IGM in gray. Note that the black curve has been multiplied by a factor of $25$ to improve visibility.

Clouds that arise from the central galaxy are shown in the left panel of Fig.~\ref{fig:app_cloudOriginHighZ}. A very small fraction of tracers were in the galaxy in the past ($\lesssim$\,$2$\,$\%$\footnote{Although if one were to consider clouds that left the galaxy only one snapshot prior, this number is $\sim$\,$85$\,$\%$.}). Although counter-intuitive, this simply reflects the rapid cycling of matter in and out of clouds, and that the present gas content of a cloud usually carries little to no information about its past \citep[see also][]{nelson2020}. Owing to poor time resolution at snapshots beyond $z$\,$\sim$\,$0.15$, the fraction of mass from the galaxy may be underestimated, since strong recycling could rapidly remove accreted gas from the central, biasing this contribution to instead be counted in the satellite and IGM categories. In either case, this suggests that the recycling of gas from the galaxy into the CGM predominantly took place $\gtrsim$\,$2$\,Gyr ago \citep{ford2016}.

On short time scales ($\lesssim$\,$500$\,Myr), a large fraction of mass was either in the cold CGM, i.e. progenitor clouds, or in the warm-hot phase of the CGM. At t$_{\rm{lookback}}$\,$\gtrsim$\,$1$\,Gyr, most of the mass is extra-galactic, after which it is smoothly accreted from the IGM ($\sim$\,$85$\,$\%$), or in a clumpy fashion ($\sim$\,$15$\,$\%$). We note that $\sim$\,$14$\,$\%$ of mass in the $z$\,$=$\,$0$ central galaxy is in satellites at $z$\,$\sim$\,$2$, while the remaining $\sim$\,$86$\,$\%$ is in the IGM. For the $z$\,$=$\,$0$ warm-hot CGM, these numbers are $\sim$\,$13$\, and $87$\,$\%$ respectively. Thus, (a) satellites contribute a slightly larger fraction to cold seeds (Fig.~\ref{fig:seedOrigin}) and a fraction of $z$\,$=$\,$0$ clouds (left- and centre- panels of Fig.~\ref{fig:app_cloudOriginHighZ}) than to the central galaxy and the warm-hot CGM, (b) the IGM contributes a smaller fraction to the central than to the warm-hot CGM, suggesting that only a fraction of material from the IGM eventually collapses into the central, while the rest remains in the CGM due to pressure support, and (c) satellites contribute a slightly larger fraction to the central than to the warm-hot gas, implying that not all satellite gas is stripped during the passage through the CGM.

In the centre-panel, we examine clouds that arise through the precipitation of the hot halo. Numbers are largely similar to the left-panel, with the main exception that a greater fraction of their $z$\,$=$\,$0$ mass is in the warm-hot CGM $\sim$\,$500$\,Myr ago. The mass fraction from the central galaxy is $\lesssim$\,$1$\,$\%$, reflecting the fraction that the cold seed acquires from the galaxy (Fig.~\ref{fig:seedOrigin}).

The right-panel focuses on clouds ejected from satellites. While a large fraction of gas is in other CGM clouds in the recent past ($\lesssim$\,$200$\,Myr), i.e. clouds with which merger events took place, $\gtrsim$\,$10$\,$\%$ of the current gas can be traced back to satellites. The inner core of these clouds is thus retained, while gas in the outer regions of the cloud is recycled. This is in contrast to clouds arising from the galaxy, in which case a greater amount of gas from the core is cycled out. We speculate that this is due to clouds ejected from satellites typically being denser (top-left panel of Fig.~\ref{fig:origin_vs_prop}), thereby allowing them to retain their cores for a longer time. Lastly, $\sim$\,$80$\,$\%$ of mass of clouds in this category is in the IGM at $z$\,$\sim$\,$2$, consistent with the scenario that IGM gas is first accreted into these satellites when they were centrals of their own halos, prior to their infall.

\section{Impact of Limited Time Resolution}\label{app:time_resl}

In this section, we assess the impact of (limited) time spacing between snapshots on a subset of results related to merger events discussed above in the main text. For this, we make use of a GIBLE halo for which we have saved a hundred snapshots between $z$\,$=$\,$0$ and $z$\,$\sim$\,$0.01$ (average snapshot spacing of $\sim$\,$1.4$\,Myr) and two hundred between $z$\,$\sim$\,$0.01$ and $z$\,$\sim$\,$0.1$ (average spacing of $\sim$\,$6.0$\,Myr). These snapshots do not contain Monte Carlo tracers that we utilise in the main work, and hence we here instead use the ParticleIDs of gas cells to construct cloud merger trees. We have verified that this is a good alternative when the time cadence is as good as $\lesssim$\,$6.0$\,Myr, particularly for those clouds that contain $\gtrsim$\,50 gas cells. Since one can neither compute mass fluxes in and out of clouds without Monte Carlo tracers, nor unambiguously assign origin channels, we restrict the following analysis to merger and fragmentation events.

\begin{figure}
    \centering
    \includegraphics[width=8cm]{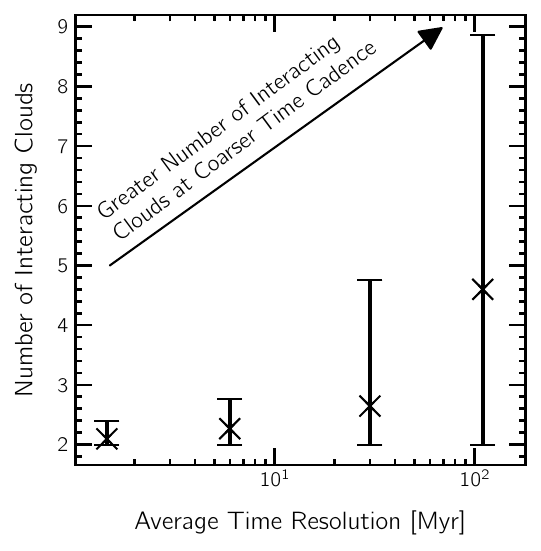}
    \caption{Number of clouds that interact during a merger event between two neighbouring snapshots (y-axis) as a function of time cadence (x-axis). Crosses show the mean, while error bars correspond to the 1$\sigma$ variation. While the average is $\sim$\,$5$ when the snapshot cadence is as coarse as $\sim$\,$100$\,Myr, it is close to $2$ at out best available snapshot spacing, suggesting that all pairwise merger events are correctly identified.}
    \label{fig:app_cldNumIntrMerg}
\end{figure}

As mentioned in Section~\ref{sssec:merg_frag}, between two snapshots, a cloud may fragment into multiple objects, or multiple clouds may merge together, or both a fragmentation and merger could take place. We therefore check the prevalence of this issue in limited time cadence snapshots. To do so, we select $z$\,$=$\,$0$ long lived clouds, i.e. those $z$\,$=$\,$0$ clouds that were already assembled $\sim$\,$2.0$\,Gyr ago, with masses in the range [$10^{5.0}$, $10^{5.4}$]\,M$_\odot$. Fig.~\ref{fig:app_cldNumIntrMerg} shows the number of clouds that interact during a merger event between two neighbouring snapshots, as a function of snapshot time cadence. Crosses show the mean, while error bars correspond to the 1$\sigma$ variation. While the average is quite high ($\sim$\,$5$) when the snapshot cadence is as coarse as $\sim$\,$100$\,Myr, at the best available time resolution ($\sim$\,$1.4$\,Myr), the average number is close to 2, i.e. events are pairwise (binary) merger between clouds. Higher numbers at worse time resolution are indicative of multiple binary mergers occuring between two neighbouring snapshots. Though not shown explicitly, a similar trend holds for fragmentation events as well.

\begin{figure}
    \centering
    \includegraphics[width=8.35cm]{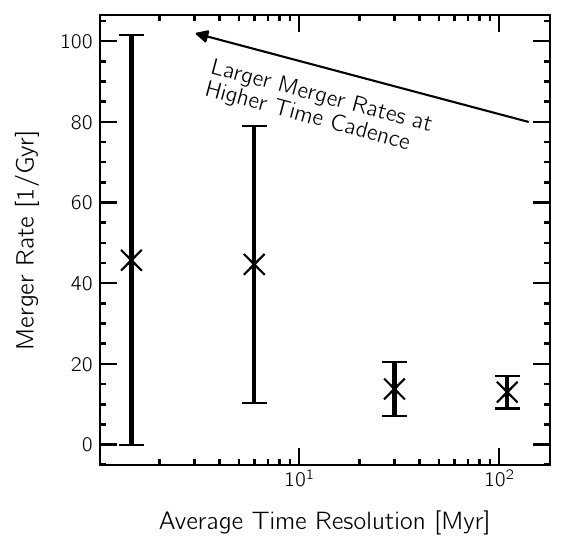}
    \caption{Merger rate of clouds (y-axis) as a function of snapshot time resolution (x-axis). Crosses correspond to the mean, while error bars show the 1$\sigma$ variation. Merger rates at coarse time resolution are underestimated, and approach a converged value at snapshot spacings of $\lesssim$\,$6.0$\,Myr.}
    \label{fig:app_mergRateTimeResl}
\end{figure}

Similarly, limited snapshot time cadence also has an impact on the computed merger rates between clouds. In Fig.~\ref{fig:app_mergRateTimeResl}, we show the merger rate as a function of the average time resolution, for the same selection of clouds as above. Crosses again show the mean, and error bars correspond to the 1$\sigma$ variation. An average rate of $\sim$\,14\,Gyr$^{-1}$ (11\,Gyr$^{-1}$) at a snapshot spacing of $\sim$\,$30$\,Myr ($100$\,Myr) is clearly an underestimate, and average merger rates as high as $\sim$\,40\,Gyr$^{-1}$ can be seen at improved time resolutions. Interestingly, the average merger rate is roughly consistent for snapshot spacings of $\sim$\,$1.0$ and $6.0$\,Myr, suggesting that these numbers reflect a converged value, although the 1$\sigma$ variation continues to increase. Resolving collective cloud dynamics in time thus requires snapshot cadence at least as good as a few Myr.

\end{document}